\documentclass[draftcls,onecolumn]{IEEEtran}
\usepackage[utf8]{inputenc}
\usepackage{amsfonts}
\usepackage{amsmath}
\usepackage{amssymb}
\usepackage{amsthm}
\usepackage[english]{babel}
\usepackage{comment}
\usepackage{xcolor}
\usepackage{graphicx}

\newtheorem{theorem}{Theorem}
\newtheorem{lemma}{Lemma}
\newtheorem{definition}{Definition}

\newtheorem*{example}{Example}
\newtheorem*{note}{Note}

\newcommand{\inv}{{\mathrm{inv}}}
\newcommand{\RANK}{\mathrm{rank}}
\newcommand{\textL}{\mathrm{L}}

\newcommand{\hr}{{\hat{r}}}

\newcommand{\MA}{{\mathbb{A}}}
\newcommand{\MB}{{\mathbb{B}}}
\newcommand{\MI}{{\mathbb{I}}}
\newcommand{\MM}{{\mathbb{M}}}
\newcommand{\MH}{{\mathbb{H}}}
\newcommand{\MO}{{\mathbb{O}}}
\newcommand{\ML}{{\mathbb{L}}}
\newcommand{\MR}{{\mathbb{R}}}
\newcommand{\MT}{{\mathbb{T}}}

\newcommand{\CR}{{\mathcal{R}}}
\newcommand{\CL}{{\mathcal{L}}}
\newcommand{\CF}{{\mathcal{F}}}
\newcommand{\CP}{{\mathcal{P}}}
\newcommand{\CI}{{\mathcal{I}}}
\newcommand{\CJ}{{\mathcal{J}}}
\newcommand{\CM}{{\mathcal{M}}}
\newcommand{\CH}{{\mathcal{H}}}

\title{Lower Bound on the Optimal Access Bandwidth of ($K+2,K,2$)-MDS Array Code with Degraded Read Friendly} 
\author{
    \IEEEauthorblockN{
        Ting-Yi Wu\IEEEauthorrefmark{1},
        Yunghsiang S.~Han\IEEEauthorrefmark{2}\IEEEauthorrefmark{1},
        Zhengrui Li\IEEEauthorrefmark{3}\IEEEauthorrefmark{1},
        Bo Bai\IEEEauthorrefmark{1},
        Gong Zhang\IEEEauthorrefmark{1},
        Liang Chen\IEEEauthorrefmark{1},
        Xiang Wu\IEEEauthorrefmark{1}
    }
    \IEEEauthorblockA{\IEEEauthorrefmark{1}%
        Huawei Technologies Co., Ltd.}\\
    \IEEEauthorblockA{\IEEEauthorrefmark{2}%
        Dongguan University of Technology}\\
    \IEEEauthorblockA{\IEEEauthorrefmark{3}%
        University of Science and Technology of China}
}

\begin{document}

\maketitle

\section{Preliminary}
\emph{Notation:} 
\begin{itemize}
 \item $\CF_q$ denotes the field of size $q$.
 
 \item $\emptyset$  denotes empty set. 
    
 \item $[N]\triangleq \{1,2,3,\ldots, N\}$ and $[0]\triangleq \emptyset$.
 
 \item $\MI\triangleq\begin{bmatrix}1 & 0\\ 0 &1\end{bmatrix}$.
 
  \item $\MO\triangleq\begin{bmatrix}0 & 0\\ 0 &0\end{bmatrix}$.
  
  \item $\CR\triangleq\left\{\begin{bmatrix}\beta_1 & 0 \\ \beta_2 & 0\end{bmatrix}: \beta_i\in\CF_{q} \mbox{ for all }i\in[2]\mbox{ and } \beta_1\beta_2\neq 0\right\}$.
  
  \item $\CL\triangleq\left\{\begin{bmatrix}0 & \beta_1 \\ 0 & \beta_2\end{bmatrix}: \beta_i\in\CF_{q} \mbox{ for all }i\in[2]\mbox{ and } \beta_1\beta_2\neq 0\right\}$.
  
  \item $\CM\triangleq\left\{\MB=\begin{bmatrix}\beta_1&\beta_2\\\beta_3&\beta_4\end{bmatrix}\not\in\{\CL\cup\CR\cup\MO\}: \beta_i\in\CF_q\mbox{ for all }i\in[4]\right\}$
  
   \item $\CM_\inv\triangleq\left\{\MB=\begin{bmatrix}\beta_1&\beta_2\\\beta_3&\beta_4\end{bmatrix}\in\CM\mbox{, and }\MB\mbox{ is invertible}: \beta_i\in\CF_q\mbox{ for all }i\in[4]\right\}$
\end{itemize}

We let the parity-check matrix of the ($K+2,K$)-MDS array be 
\begin{equation}\label{eq:upperI_H}
            \MH=\begin{bmatrix}\MI & \MI &\cdots & \MI \\ \MA_1 & \MA_2 & \cdots &\MA_N\end{bmatrix},
\end{equation}
where $N=K+2$ and these upper identities make the degraded read friendly. 
To satisfy MDS property,  $\begin{bmatrix}\MI & \MI\\\MA_i & \MA_j\end{bmatrix}$ must be invertible for any $i\neq  j$, 
which is equivalent to that  $(\MA_i+\MA_j)$ must be invertible \cite{Silvester2000determinants} for any $i\neq  j$, 
For any node $i\in [N]$, it stores two symbols as a vector $\alpha^{(i)}\triangleq\begin{bmatrix}\alpha^{(i)}_1\\ \alpha^{(i)}_2\end{bmatrix}$.

To repair the $i$th node, we have to find two check equations from the row space of  $\MH$ in order to retrieve $\alpha^{(i)}$. 
Let $\MM$ be a repair matrix of size $2\times 4$, $\MM\MH$ denotes two check equations from the row space of $\MH$ such that 
\begin{equation}
    \MM\MH\triangleq\begin{bmatrix}\MH_1&\MH_2&\cdots&\MH_N\end{bmatrix},
\end{equation}
and
\begin{equation}
    \MH_1\alpha^{(1)}+\MH_2\alpha^{(2)}+\cdots +\MH_N\alpha^{(N)}=\begin{bmatrix}0\\0\end{bmatrix},
\end{equation}
where $\MH_i=\MM\begin{bmatrix}\MI\\\MA_i\end{bmatrix}$ for all $i\in[N]$.
If $\MM$ can be used to recover $\alpha^{(i)}$, we have to make sure that $\MH_i$ is invertible and recover $\alpha^{(i)}$ as
\begin{equation}
    \alpha^{(i)}=\MH_i^{-1}\left[\sum_{j\in[N]\setminus\{i\}}\MH_j\alpha^{(j)}\right].
\end{equation}

To be able to repair each node, we have to design the corresponding repair matrix for each $\alpha^{(i)}$. 
Since some repair matrices can be used to recover multiple nodes, i.e., $\MM\begin{bmatrix}\MI\\\MA_i\end{bmatrix}$ and $\MM\begin{bmatrix}\MI\\\MA_j\end{bmatrix}$ are both invertible for some $i\neq j$, we are not restricted to find $N$ different repair matrices for recovering all $N$ nodes. 
Let $R$ be the number of repair matrices, $1\leq R\leq N$, and $\{\CI^{(1)}, \CI^{(2)}, \ldots, \CI^{(R)}\}$ be a partition of $[N]$,  
a repair process of size $R$ is formed by $R$ repair matrices, $\MM^{(1)},\MM^{(2)}, \ldots,\MM^{(R)}$, where $\MM^{(r)}$ can be used to recover $\alpha^{(i)}$ is $i\in\CI^{(r)}$. 
The formal definition of a repiar process of size $R$ is given below. 

\begin{definition}\label{def:repair_process}
    A repair process $P_R$ of $R$ repair matrices is defined as 
    \begin{equation}
        P_R\triangleq\left\{\left(\MM^{(r)},\CI^{(r)}\right): r\in[R] \right\},
    \end{equation}
    such that $\CI^{(1)}$, $\CI^{(2)}$, $\ldots$, $\CI^{(R)}$ form a partition of $[N]$ and $\MM^{(r)}\begin{bmatrix}\MI\\ \MA_i \end{bmatrix}$ is invertible for $i\in \CI^{(r)}$.
\end{definition}
We then explain the process of recovering $\alpha^{(i)}$ by a $P_R$, where $i\in\CI^{(r)}$. 
Let 
\begin{equation}
    \MM^{(r)}\MH\triangleq\begin{bmatrix}\MH^{(r)}_1 & \MH^{(r)}_2 & \cdots & \MH^{(r)}_N\end{bmatrix}, 
\end{equation}
where $\MH^{(r)}_i=\MM^{(r)}\begin{bmatrix}\MI\\\MA_i\end{bmatrix}$.
 By Definition \ref{def:repair_process}, $\MH^{(r)}_i$ is invertible, therefore 
\begin{equation}
    \alpha^{(i)}=\left[\MH^{(r)}_i\right]^{-1} \left[\sum_{j\in[N]\setminus\{i\}} \MH^{(r)}_j \alpha^{(j)}\right].
\end{equation}

\begin{definition}\label{def:repair_bandwidth}
 The repair bandwidth for repairing node $i$, where $i\in\CI^{(r)}$, by using $P_R$ is denoted by 
 \begin{equation}
     B_i(P_R)=\sum_{j\in [N]\setminus \{i\}} B_{i,j}(P_R),
 \end{equation}
 and
 \begin{equation}\label{eq:Bij}
     B_{i,j}(P_R)=
     \begin{cases}
         0, & \mbox{if } i=j \mbox{ or }\MH^{(r)}_j=\MO;\\
         1, & \mbox{if } \MH^{(r)}_j\in\CL \mbox{ or } \MH^{(r)}_j\in\CR;\\
         2, & \mbox{otherwise.}
     \end{cases}
 \end{equation}
\end{definition}

 Note that $B_{i,j}(P_R)$ denotes the number of symbols needed to be download from $j$ to repair node $i$ by using repair process $P_R$. 
 
 \vspace{1cm}
 \begin{lemma}
     Given a $P_R$, $B_{i,j}(P_R)=B_{j,i}(P_R)=2$ if $i\neq  j$ and $i,j\in\CI^{(r)}$ for  $r\in [R]$.
 \end{lemma}\label{lm:Bij2}
 \begin{IEEEproof}
     By Definition \ref{def:repair_process}, $\MH^{(r)}_i$ and $\MH^{(r)}_j$ are both invertible since $i,j\in\CI^{(r)}$. 
         Therefore, $B_{i,j}(P_R)=B_{j,i}(P_R)=2$ by \eqref{eq:Bij}. 
 \end{IEEEproof}

 \vspace{1cm}
 \begin{lemma}\label{lm:sameBiBj}
     Given a $P_R$, $B_i(P_R)=B_j(P_R)$ when both $i,j\in\CI^{(r)}$ for some $r$.
 \end{lemma}
 \begin{IEEEproof}
     By Definition \ref{def:repair_bandwidth}, $B_i(P_R)=\sum_{k\in[N]\setminus\{i\}}B_{i,k}(P_R)$ and $B_{i,k}(P_R)=B_{j,k}(P_R)$ for all $k\in[N]\setminus\{i,j\}$.
    Since $B_{i,j}(P_R)=B_{j,i}(P_R)$, we have
 \begin{IEEEeqnarray}{rCl}
     B_i(P_R)&=&\sum_{k\in[N]\setminus\{i\}}B_{i,k}(P_R)\\
    &=&\sum_{k\in[N]\setminus\{i,j\}}B_{i,k}(P_R)+B_{i,j}(P_R)\\
    &=&\sum_{k\in[N]\setminus\{i,j\}}B_{j,k}(P_R)+B_{j,i}(P_R)\\
    &=&\sum_{k\in[N]\setminus\{j\}}B_{j,k}(P_R)=B_j(P_R).
 \end{IEEEeqnarray}
 \end{IEEEproof}

 From Lemma \ref{lm:sameBiBj}, we have $B_i(P_R)=B_j(P_R)$ for all $i,j\in\CI^{(r)}$ and $i\neq j$.
 Hence, we define $B^{(r)}(P_R)$ as below. 
 \begin{definition}\label{def:Br}
     Let $B^{(r)}(P_R)$ be the repair bandwidth for repairing the node in $\CI^{(r)}$, i.e., $B^{(r)}(P_R)= B_i(P_R)$ for all $i\in\CI^{(r)}$. 
 \end{definition}
 
 \vspace{1cm}
 \begin{lemma}\label{lm:totalrepairbandwidth}
     The total repair bandwidth of $P_R$ is $B(P_R)\triangleq\sum_{i\in[N]}B_i(P_R)=\sum_{r\in[R]}|\CI^{(r)}|B^{(r)}(P_R)$, where $|\CI^{(r)}|$ is the number of distinct elements in set $\CI^{(r)}$.
 \end{lemma}
 \begin{IEEEproof}
     By Definition \ref{def:Br}, 
     \begin{IEEEeqnarray}{rCl}
         \sum_{i\in[N]}B_i(P_R)&=&\sum_{r\in[R]}\sum_{i\in\CI^{(r)}}B_i(P_R)\\
         &=&\sum_{r\in[R]}\sum_{i\in\CI^{(r)}}B^{(r)}(P_R)\\
         &=&\sum_{r\in[R]}|\CI^{(r)}|B^{(r)}(P_R).\\
     \end{IEEEeqnarray}
 \end{IEEEproof}

\begin{example}
    \begin{itemize}
        \item $\MH=\begin{bmatrix}
            1  & 0 & 1 & 0&  1&  0\\0  &1 &  0 & 1&  0&  1\\2  &0&  0 & 4&  0&  8\\0  &3&  5 &  1&  9&  1 \end{bmatrix}$, where $N=3$.
        
        \item $P_2=\left(\MM^{(r)},\CI^{(r)}\right)_{r\in[2]}$, where 
        
        $\MM^{(1)}=\begin{bmatrix}1&0&0&0\\0&1&0&1\end{bmatrix}$ and $\CI^{(1)}=\{1\}$, 
        
        $\MM^{(2)}=\begin{bmatrix}0&1&0&0\\0&0&0&1\end{bmatrix}$ and $\CI^{(2)}=\{2,3\}$.
        
        \item $\MM^{(1)}\MH= \begin{bmatrix}\MH^{(1)}_1 & \MH^{(1)}_2 &  \MH^{(1)}_3\end{bmatrix}=\begin{bmatrix}1&0&1&0&1&0\\0&2&5&0&9&0\end{bmatrix}\triangleq\begin{bmatrix}\MB&\MR_1&\MR_2\end{bmatrix}$, where $\MB\in\CM_\inv$ and $\MR_i\in\CR$ for all $i\in[2]$. 

        $\MM^{(2)}\MH=\begin{bmatrix}\MH^{(2)}_1 & \MH^{(2)}_2 &  \MH^{(2)}_3\end{bmatrix}=\begin{bmatrix}0&1&0&1&0&1\\0&3&5&1&9&1\end{bmatrix}\triangleq\begin{bmatrix}\ML&\MB_1&\MB_2\end{bmatrix}$, where $\MB_i\in\CM_\inv$ for all $i\in[2]$ and $\ML\in\CR$. 
        
        \item Repairing node 1:
        
        \begin{IEEEeqnarray}{rCl}
            \MH^{(1)}_1\alpha^{(1)}+\MH^{(1)}_2\alpha^{(2)}+\MH^{(1)}_3\alpha^{(3)}=\begin{bmatrix}0\\0\end{bmatrix}
            &\Rightarrow&\alpha^{(1)}=\begin{bmatrix}\MH^{(1)}_1\end{bmatrix}^{-1}
            \left(\MH^{(1)}_2\alpha^{(2)}+\MH^{(1)}_3\alpha^{(3)}\right)\\
            &\Rightarrow&\alpha^{(1)}=\begin{bmatrix}1&0\\0&2\end{bmatrix}^{-1}\left(\begin{bmatrix}1&0\\5&0\end{bmatrix}\alpha^{(2)}+\begin{bmatrix}1&0\\9&0\end{bmatrix}\alpha^{(3)}\right).                
        \end{IEEEeqnarray}
        
         \item Repairing node 2:
         
          \begin{IEEEeqnarray}{rCl}
            \MH^{(2)}_1\alpha^{(1)}+\MH^{(2)}_2\alpha^{(2)}+\MH^{(2)}_3\alpha^{(3)}=\begin{bmatrix}0\\0\end{bmatrix}
            &\Rightarrow&\alpha^{(2)}=\begin{bmatrix}\MH^{(2)}_2\end{bmatrix}^{-1}
            \left(\MH^{(2)}_1\alpha^{(1)}+\MH^{(2)}_3\alpha^{(3)}\right)\\
            &\Rightarrow&\alpha^{(2)}=\begin{bmatrix}0&1\\5&1\end{bmatrix}^{-1}\left(\begin{bmatrix}0&1\\0&3\end{bmatrix}\alpha^{(2)}+\begin{bmatrix}0&1\\9&1\end{bmatrix}\alpha^{(3)}\right).                
        \end{IEEEeqnarray}
        
         \item Repairing node 3:
         
          \begin{IEEEeqnarray}{rCl}
            \MH^{(2)}_1\alpha^{(1)}+\MH^{(2)}_2\alpha^{(2)}+\MH^{(2)}_3\alpha^{(3)}=\begin{bmatrix}0\\0\end{bmatrix}
            &\Rightarrow&\alpha^{(3)}=\begin{bmatrix}\MH^{(2)}_3\end{bmatrix}^{-1}
            \left(\MH^{(2)}_1\alpha^{(1)}+\MH^{(2)}_2\alpha^{(2)}\right)\\
            &\Rightarrow&\alpha^{(2)}=\begin{bmatrix}0&1\\9&1\end{bmatrix}^{-1}\left(\begin{bmatrix}0&1\\0&3\end{bmatrix}\alpha^{(1)}+\begin{bmatrix}0&1\\5&1\end{bmatrix}\alpha^{(2)}\right).                
        \end{IEEEeqnarray}
        
        \item $B_{i,j}(P_R)=\begin{bmatrix}0&1&1\\1&0&2\\1&2&0\end{bmatrix}$, $B_1(P_R)=2$, $B_2(P_R)=3$, and $B_3(P_R)=3$.
    \end{itemize}
\end{example}

\section{Some properties}
\begin{lemma}\label{lm:allzeros}
    Given a $P_R$ and $r\in[R]$,
    if there is a $B_{i,j}(P_R)=0$  for some $i\in\CI^{(r)}$ and $j\in[N]\setminus\CI^{(r)}$, then $B^{(r)}(P_R)=B_i(P_R)=2K$.
\end{lemma}
\begin{IEEEproof}
    Since $i\neq j$, $B_{i,j}(P_R)=0$ implies $\MH^{(r)}_j=\MO$ due to $\eqref{eq:Bij}$.
    Let $\MM^{(r)}=\begin{bmatrix}\MM^{(r)}_1 & \MM^{(r)}_2\end{bmatrix}$ be a repair matrix of $P_R$, we have an invertible matrix $\MH^{(r)}_i$ such that
    \begin{multline}\label{eq:invertible_sum}
        \MH^{(r)}_i=\MH^{(r)}_i+\MH^{(r)}_j=\MM^{(r)}\begin{bmatrix}\MI\\\MA_i\end{bmatrix}+\MM^{(r)}\begin{bmatrix}\MI\\\MA_j\end{bmatrix}=\MM^{(r)}\begin{bmatrix}\MI+\MI\\\MA_i+\MA_j\end{bmatrix}\\
        =\begin{bmatrix}\MM^{(r)}_1&\MM^{(r)}_2\end{bmatrix}\begin{bmatrix}\MO\\\MA_i+\MA_j\end{bmatrix}
        =\MM^{(r)}_1\MO+\MM^{(r)}_2(\MA_i+\MA_j)=\MM^{(r)}_2(\MA_i+\MA_j).
    \end{multline}
    
     Since $\MH^{(r)}_i$ and $(\MA_i+\MA_j)$ are invertible, we conclude that $\MM^{(r)}_2$ is also invertible.
     From \eqref{eq:invertible_sum}, we have $\MH^{(r)}_k=\MM^{(r)}_2(\MA_k+\MA_j)$ for all $k\in[N]$. Since both $\MM^{(r)}_2$ and $(\MA_k+\MA_j)$ are invertible if $k\neq  j$, $\MH^{(r)}_k$ is invertible if $k\neq j$. 
     Therefore, from \eqref{eq:Bij}, we have $B^{(r)}(P_R)=B_{i}(P_R)=\sum_{k\in[N]\setminus\{i\}}B_{i,k}(P_R)=\sum_{k\in[N]\setminus\{i,j\}}2=2(N-2)=2K$.
\end{IEEEproof}
\begin{example}
    $$\MM^{(r)}\MH=\begin{bmatrix}
    1&0&1&0&1&0&0&0\\
    0&1&2&3&4&5&0&0
\end{bmatrix}\triangleq\begin{bmatrix}\MB_1&\MB_2&\MB_3&\MO\end{bmatrix}$$
is okay, but 
$$\MM^{(r)}\MH=\begin{bmatrix}
    1&0&1&0&1&0&0&0\\
    0&1&2&3&4&0&0&0
\end{bmatrix}\triangleq\begin{bmatrix}\MB_4&\MB_5&\MR&\MO\end{bmatrix}$$
is impossible, where $\MR\in\CR$ and $\MB_i\in\CM_\inv$ for all $i\in[5]$.
\end{example}

\vspace{1cm}
\begin{lemma}\label{lm:fullrankM}
    Given a $P_R$, then $\MM^{(r)}$ is of rank $2$ for all $r\in[R]$. 
\end{lemma}
 \begin{note}
     This lemma can be applied to $N-K\geq 2$.
    \end{note}
\begin{IEEEproof}
    Since $\RANK(\MA\MB)\leq\min\{\RANK(\MA),\RANK(\MB)\}$ and $\MM^{(r)}\begin{bmatrix}\MI\\\MA_i\end{bmatrix}$ is invertible if $i\in\CI^{(r)}$,  
    \begin{IEEEeqnarray}{rCl}
        \RANK\left(\MM^{(r)}\begin{bmatrix}\MI\\\MA_i\end{bmatrix}\right)=2&\leq&
        \min\left\{\RANK(\MM^{(r)}), \RANK\left(\begin{bmatrix}\MI\\\MA_i\end{bmatrix}\right)\right\}\\
        &\leq&\RANK(\MM^{(r)})\\
        &\leq&2.
    \end{IEEEeqnarray}
    Therefore, $\RANK(\MM^{(r)})=2$ for all $r\in[R]$.
\end{IEEEproof}

\vspace{1cm}
\begin{lemma}\label{lm:rankM1M2}
    Given a $P_R$,  if $\MH^{(r_1)}_i$ and  $\MH^{(r_2)}_i$ are both in  $\CL$ (or $\CR$),  $\MH^{(r_1)}_j$ and  $\MH^{(r_2)}_j$
    are both in $\CL$ (or $\CR$) for some $r_1,r_2\in[R]$, $i,j\in[N]\setminus\{\CI^{(r_1)}\cup\CI^{(r_2)}\}$, and $i\neq j$, then $\begin{bmatrix}\MM^{(r_1)}\\\MM^{(r_2)}\end{bmatrix}$ is of rank $2$.
\end{lemma}
   \begin{note}
     This lemma can be applied to $N-K\geq 2$.
    \end{note}
\begin{IEEEproof}
    Since $\begin{bmatrix}\MI&\MI\\\MA_i&\MA_j\end{bmatrix}$ is invertible and $\RANK(\MA)=\RANK(\MA\MB)$ if $\MB$ is of full rank,
    \begin{equation}\label{eq:rankM1M2}
    \RANK\left(\begin{bmatrix}\MM^{(r_1)}\\\MM^{(r_2)}\end{bmatrix}\right)=
    \RANK\left(\begin{bmatrix}\MM^{(r_1)}\\\MM^{(r_2)}\end{bmatrix}
    \begin{bmatrix}\MI&\MI\\\MA_i&\MA_j\end{bmatrix}\right)
    =\RANK\left(\begin{bmatrix}\MH^{(r_1)}_i & \MH^{(r_2)}_j\\\MH^{(r_2)}_i&\MH^{(r_2)}_j\end{bmatrix}\right)\leq 2.
\end{equation}
Combining \eqref{eq:rankM1M2} with Lemma \ref{lm:fullrankM}, we conclude that 
$\begin{bmatrix}\MM^{(r_1)}\\\MM^{(r_2)}\end{bmatrix}$ is of rank $2$ for all $r_1,r_2\in [R]$.
\end{IEEEproof}

\vspace{1cm}
\begin{lemma}\label{lm:equivM}
    If $\MM^{(r_1)}=\MT\MM^{(r_2)}$ for some $r_1,r_2\in[R]$, $r_1\neq r_2$, and invertible matrix $\MT$, then both $\MM^{(r_1)}$ and $\MM^{(r_2)}$ can be used to repair nodes in $\CI^{(r_1)}\cup\CI^{(r_2)}$ and $B^{(r_1)}(P_R)=B^{(r_2)}(P_R)$.
\end{lemma}
\begin{IEEEproof}
    Since $\MT$ is invertible, $\MH^{(r_2)}_i$ is invertible if and only if  $\MH^{(r_1)}_i=\MT\MH^{(r_2)}_i$ is invertible. 
    With the facts that $\MH^{(r_1)}_i$ is invertible when $i\in\CI^{(r_1)}$ and $\MH^{(r_2)}_j$ is invertible when $j\in\CI^{(r_2)}$, we can conclude that both $\MH^{(r_1)}_k$ and $\MH^{(r_2)}_k$ are invertible when $k\in\CI^{(r_1)}\cup\CI^{(r_2)}$. 
    Therefore, both $\MH^{(r_1)}$ and $\MH^{(r_2)}$ can be used to repair nodes in $\CI^{(r_1)}\cup\CI^{(r_2)}$. 
   
   From \eqref{eq:Bij}, $B_{i,j}(P_R)=1$, for some $i\in\CI^{(r_2)}$, denotes that $\MH^{(r_2)}_j\in\{\CL\cup\CR\}$, so $\MH^{(r_1)}_j=\MT\MH^{(r_2)}_j\in\{\CL\cup\CR\}$. 
   Also, $B_{i,j}(P_R)=1$, for some $i\in\CI^{(r_1)}$, denotes that $\MH^{(r_1)}_j\in\{\CL\cup\CR\}$, so $\MH^{(r_2)}_j=\MT^{-1}\MM^{(r_1)}_j\in\{\CL\cup\CR\}$.
   
   Furthermore, $B_{i,j}(P_R)=0$, for some $i\in\CI^{(r_2)}$ and $i\neq j$, denotes that $\MH^{(r_2)}_j=\MO$, so $\MH^{(r_1)}_j=\MT\MH^{(r_2)}_j=\MO$. 
   Also, $B_{i,j}(P_R)=0$, for some $i\in\CI^{(r_1)}$ and $i\neq j$, denotes that $\MH^{(r_1)}_j=\MO$, so $\MH^{(r_2)}_j=\MT^{-1}\MM^{(r_1)}_j=\MO$.
   
  We then prove that if $B_{i,j}(P_R)=2$ for some $i\in\CI^{(r_1)}$ and $i\neq j$, then $\MH^{(r_2)}_j\not\in\{\MO\cup\CR\cup\CL\}$. 
  Assuming $\MH^{(r_2)}_j\in\{\MO\cup\CR\cup\CL\}$, so $\MH^{(r_1)}_j=\MT^{-1}\MH^{(r_2)}_j\in\{\MO\cup\CR\cup\CL\}$. 
  Hence we can conclude that $B_{i,j}(P_R)\neq 2$ from \eqref{eq:Bij}.
  By contradiction, $B_{i,j}(P_R)=2$, for some $i\in\CI^{(r_1)}$ and $i\neq j$, implies $\MH^{(r_2)}_j\not\in\{\MO\cup\CR\cup\CL\}$.
  Following the similar way, we can also prove that if $B_{i,j}(P_R)=2$ for some $i\in\CI^{(r_2)}$ and $i\neq j$, then $\MH^{(r_1)}_j\not\in\{\MO\cup\CR\cup\CL\}$. 
  
  Combining above results and \eqref{eq:Bij}, we can conclude that $B_{i,k}(P_R)=B_{j,k}(P_R)$ for some $i\in\CI^{(r_1)}$, $j\in\CI^{(r_2)}$, and for all $k\in[N]\setminus\{i,j\}$. Therefore, $B^{(r_1)}(P_R)=B^{(r_2)}(P_R)$.
\end{IEEEproof}

\vspace{1cm}
\begin{theorem}\label{thm:girth4free}
    For any $P_R$ such that  $\MH^{(r_1)}_i$ and  $\MH^{(r_2)}_i$ are both in  $\CL$ (or $\CR$),  $\MH^{(r_1)}_j$ and  $\MH^{(r_2)}_j$
    are both in $\CL$ (or $\CR$) for some $r_1,r_2\in[R]$, and $i,j\in[N]\setminus\{\CI^{(r_1)}\cup\CI^{(r_2)}\}$, where  $r_1\neq r_2$ and $i\neq j$, 
    then the repair process
    \begin{equation}P_{R-1}=\left\{\left(\MM^{(r)},\CI^{(r)}\right):r\in[R]\setminus\{r_1,r_2\}\right\}\cup\left\{\left(\MM^{(r_1)},\CI^{(r_1)}\cup\CI^{(r_2)}\right)\right\}
    \end{equation}
 has the same total repair bandwidth as $P_R$ has. 
\end{theorem}
\begin{IEEEproof}
    By Lemmas \ref{lm:fullrankM} and \ref{lm:rankM1M2}, we can have an invertible transform matrix $\MT$ such that $\MM^{(r_1)}=\MT\MM^{(r_2)}$.
    Then according to Lemma \ref{lm:equivM}, $\MM^{(r_1)}$ can repair nodes in $\CI^{(r_2)}$ and $B^{(r_1)}(P_R)=B^{(r_2)}(P_R)$. Therefore,  
    \begin{IEEEeqnarray}{rCl}
        B(P_R)&=&\sum_{r\in[R]}|\CI^{(r)}|B^{(r)}(P_R)\\
        &=&\sum_{r\in[R]\setminus\{r_1,r_2\}}|\CI^{(r)}|B^{(r)}(P_R)+|\CI^{(r_1)}|B^{(r_1)}(P_R)+|\CI^{(r_2)}|B^{(r_2)}(P_R)\\
        &=&\sum_{r\in[R]\setminus\{r_1,r_2\}}|\CI^{(r)}|B^{(r)}(P_R)+|\CI^{(r_1)}|B^{(r_1)}(P_R)+|\CI^{(r_2)}|B^{(r_1)}(P_R)\\
        &=&\sum_{r\in[R]\setminus\{r_1,r_2\}}|\CI^{(r)}|B^{(r)}(P_R)+|\CI^{(r_1)}+\CI^{(r_1)}|B^{(r_1)}(P_R)\\
        &=&B(P_{R-1}).
    \end{IEEEeqnarray}
\end{IEEEproof}

\vspace{1cm}
\begin{lemma}\label{lm:Bicalculation}
 Given a $P_R$, if $B_i(P_R)<2K$ for some $i\in\CI^{(r)}$, then
 \begin{IEEEeqnarray}{rCl}
  B_i(P_R)&=&2(N-1)-\left|\left\{j: j\in[N]\setminus\{i\} \mbox{ and }\MH^{(r)}_j\in\{\CL\cup\CR\}\right\}\right|\\
  &=&2(N-1)-\left|\left\{j: j\in[N]\setminus\{i\} \mbox{ and }B_{i,j}(P_R)=1\right\}\right|,
 \end{IEEEeqnarray}
where $\MH^{(r)}_j=\MM^{(r)}\begin{bmatrix}\MI\\\MA_j\end{bmatrix}$.
\end{lemma}
\begin{IEEEproof}
Due to Lemma \ref{lm:allzeros} and $B_i(P_R)<2K$, we can not have $B_{i,j}(P_R)=0$ for all $j\in[N]\setminus\{i\}$.  
By Definition \ref{def:repair_bandwidth}, 
\begin{IEEEeqnarray}{rCl}
 B_i(P_R)&=&\sum_{j\in[N]\setminus\{i\}}B_{i,j}(P_R)\\
 &=&2\left|\{j: j\in[N]\setminus\{i\}\mbox{ and }\MH^{(r)}_j\in\CM\}\right|+\left|\{j: j\in[N]\setminus\{i\}\mbox{ and }\MH^{(r)}_j\in\{\CL\cup\CR\}\}\right|\\
 &=&2(N-1)-\left|\{j: j\in[N]\setminus\{i\}\mbox{ and }\MH^{(r)}_j\in\{\CL\cup\CR\}\}\right|\\
 &=&2(N-1)-\left|\left\{j: j\in[N]\setminus\{i\} \mbox{ and }B_{i,j}(P_R)=1\right\}\right|
\end{IEEEeqnarray}
\end{IEEEproof}

\vspace{1cm}
We then present the properties of repair matrices of any $P_R$. 
\begin{lemma}\label{lm:M2notinvertible}
    Given a $P_R$, 
    if $B_i(P_R)<2K$  for some $i\in\CI^{(r)}$, then $\MM^{(r)}_2$ is not invertible, where $\MM^{(r)}=\begin{bmatrix}\MM^{(r)}_1&\MM^{(r)}_2\end{bmatrix}$. 
\end{lemma}
\begin{IEEEproof}
    Since $B_i(P_R)<2K$, we have 
    \begin{equation}
     B_i(P_R)=2(N-1)-\left|\left\{j: j\in[N]\setminus\{i\} \mbox{ and }\MH^{(r)}_j\in\{\CL\cup\CR\}\right\}\right|<2K,
    \end{equation}
    which induces
    \begin{equation}
    \left|\left\{j: j\in[N]\setminus\{i\} \mbox{ and }\MH^{(r)}_j\in\{\CL\cup\CR\}\right\}\right|>2(N-1)-2K=2.
    \end{equation}
    By pigeonhole principle, there must be $j_1, j_2\in[N]\setminus\{i\}$, such that $j_1\neq j_2$ and both $\MH^{(r)}_{j_1}, \MH^{(r)}_{j_2}\in\CL$ (or $\CR$). 
    WLOG, we assume $\MH^{(r)}_{j_1},\MH^{(r)}_{j_2}\in\CL$ and then we have
    \begin{IEEEeqnarray}{rCl}
    \MH^{(r)}_{j_1}+\MH^{(r)}_{j_2}&=&
     \MM^{(r)}\begin{bmatrix}\MI\\\MA_{j_1}\end{bmatrix}+\MM^{(r)}\begin{bmatrix}\MI\\\MA_{j_2}\end{bmatrix}\\
     &=&\begin{bmatrix}\MM^{(r)}_1&\MM^{(r)}_2\end{bmatrix}\begin{bmatrix}\MO\\\MA_{j_1}+\MA_{j_2}\end{bmatrix}\\
     &=&\MM^{(r)}_2(\MA_{j_1}+\MA_{j_2})\in\CL.
    \end{IEEEeqnarray}
    Since $\MM^{(r)}_2(\MA_{j_1}+\MA_{j_2})$ is not invertible and $\MA_{j_1}+\MA_{j_2}$ is invertible, we can conclude that $\MM^{(r)}_2$ is not invertible.
\end{IEEEproof}

\vspace{1cm}
\begin{lemma}\label{lm:hatM}
 Given a $P_R=\left\{\left(\MM^{(r)},\CI^{(r)}\right): r\in[R] \right\}$ 
 and $\hr\in[R]$ 
 such that  $B_{i}(P_R)<2K$ for some ${i}\in\CI^{(\hr)}$.
 By letting $\hat{\MM}^{(\hr)}=\left[\MH^{(\hr)}_{i}\right]^{-1}\MM^{(\hr)}$,  
 \begin{description}
\item[(1)] we  can construct an repair process $\hat{P}_R$ as
 \begin{equation}
    \hat{P}_R=\left\{\left(\MM^{(r)},\CI^{(r)}\right): r\in[R]\setminus\{\hr\} \right\}\cup
    \left\{\left(\hat{\MM}^{(\hr)},\CI^{(\hr)}\right)\right\},
 \end{equation}
 where  $B_i(P_R)=B_i(\hat{P}_R)$.
\item[(2)] For $\hat{P}_R$,
 \begin{equation}
  \hat{\MM}^{(\hr)}=\begin{bmatrix}1&0&0&0\\\beta^{(\hr)}_1&\beta^{(\hr)}_2&\beta^{(\hr)}_3&\beta^{(\hr)}_4\end{bmatrix},
 \end{equation}
 if $\exists j_1,j_2\in[N]$ and $j_1\neq j_2$ such that $\hat{\MH}^{(\hr)}_{j_1},\hat{\MH}^{(\hr)}_{j_2}\in\CL$, 
 where $\hat{\MH}^{(\hr)}_{j}=
  \hat{\MM}^{(\hr)}\begin{bmatrix}\MI\\\MA_{j}\end{bmatrix}$ for all $j\in[N]$ and 
  $\beta^{(\hr)}_k\in\CF_q$ for all $k\in[4]$.
  
 \item[(3)] For $\hat{P}_R$, 
   \begin{equation}
  \hat{\MM}^{(\hr)}=\begin{bmatrix}\beta^{(\hr)}_1&\beta^{(\hr)}_2&\beta^{(\hr)}_3&\beta^{(\hr)}_4\\0&1&0&0\end{bmatrix},
 \end{equation}
  if $\exists j_1,j_2\in[N]$ and $j_1\neq j_2$ such that $\hat{\MH}^{(\hr)}_{j_1},\hat{\MH}^{(\hr)}_{j_2}\in\CR$.
  \end{description}
\end{lemma}
\begin{IEEEproof}
 (1) is followed directly from Lemma~\ref{lm:equivM}. 
 
 Next we prove (2). Since $\hat{\MM}^{(\hr)}=\left[\MH^{(\hr)}_{i}\right]^{-1}\MM^{(\hr)}$, 
 \begin{equation}
  \hat{\MH}^{(\hr)}_{i}\triangleq
  \hat{\MM}^{(\hr)}\begin{bmatrix}\MI\\\MA_{i}\end{bmatrix}
  =\left[\MH^{(\hr)}_{i}\right]^{-1}\MM^{(\hr)}\begin{bmatrix}\MI\\\MA_{i}\end{bmatrix}
  =\left[\MH^{(\hr)}_{i}\right]^{-1}\MH^{(\hr)}_{i}=\MI.
 \end{equation}
 Since $B_i(\hat{P}_R)<2K$, $\hat{\MM}^{(\hr)}_2$ is not invertible due to Lemma \ref{lm:M2notinvertible}.

 We first prove that $\hat{\MH}^{(\hr)}_j=\begin{bmatrix}1&0\\\beta&0\end{bmatrix}$  for some $\beta\in\CF_q$ if $\hat{\MH}^{(\hr)}_j\in\CL$.
 Let $\hat{\MH}^{(\hr)}_j=\begin{bmatrix}\beta'&0\\\beta&0\end{bmatrix}\in\CL$ for some $\beta,\beta'\in\CF_q$, then we have
 \begin{equation}
    \hat{\MH}^{(\hr)}_{i}+\hat{\MH}^{(\hr)}_j
    =\MI+\begin{bmatrix}\beta'&0\\\beta&0\end{bmatrix}
    =\begin{bmatrix}1+\beta'&0\\\beta&1\end{bmatrix}
    =\hat{\MM}^{(\hr)}_2(\MA_{i}+\MA_j).
 \end{equation}
 Since $\hat{\MM}^{(\hr)}_2$ is not invertible, then  
 $\begin{bmatrix}1+\beta'&0\\\beta&1\end{bmatrix}$ is not invertible as well, 
 which implies that the determinant of $\begin{bmatrix}1+\beta'&0\\\beta&1\end{bmatrix}$ is $0$. 
    Therefore, $1\times(1+\beta')=0$ implies $\beta'=1$. 

    We next prove that $\hat{\MM}^{(\hr)}_2=\begin{bmatrix}0&0\\\beta^{(\hr)}_3&\beta^{(\hr)}_4\end{bmatrix}$ for some $\beta^{(\hr)}_3,\beta^{(\hr)}_4\in\CF_q$,  
    if $\exists j_1,j_2\in[N]$ and $j_1\neq j_2$ such that $\hat{\MH}^{(\hr)}_{j_1},\hat{\MH}^{(\hr)}_{j_2}\in\CL$. 
    Let $\hat{\MH}^{(\hr)}_{j_1}=\begin{bmatrix}1&0\\\beta_1&0\end{bmatrix}$ and  $\hat{\MH}^{(\hr)}_{j_2}=\begin{bmatrix}1&0\\\beta_2&0\end{bmatrix}$, we have
    \begin{IEEEeqnarray}{rCl}
        \hat{\MH}^{(\hr)}_{j_1}+\hat{\MH}^{(\hr)}_{j_2}
        &=&\begin{bmatrix}0&0\\\beta_1+\beta_2&0\end{bmatrix}\\
        &=&\hat{\MM}^{(\hr)}_{2}(\MA_{j_1}+\MA_{j_2}).
    \end{IEEEeqnarray}
    Since $(\MA_{j_1}+\MA_{j_2})$ is invertible, then 
    \begin{equation}
     \hat{\MM}^{(\hr)}_2=\begin{bmatrix}0&0\\\beta_1+\beta_2&0\end{bmatrix}(\MA_{j_1}+\MA_{j_2})^{-1}
     =\begin{bmatrix}0&0\\\beta^{(\hr)}_3&\beta^{(\hr)}_4\end{bmatrix},
    \end{equation}
   for some $\beta^{(\hr)}_3,\beta^{(\hr)}_4\in \CF_q$.
    Lastly, we prove that $\hat{\MM}^{(\hr)}_1=\begin{bmatrix}1&0\\\beta^{(\hr)}_1&\beta^{(\hr)}_2\end{bmatrix}$ for some $\beta^{(\hr)}_1,\beta^{(\hr)}_2\in\CF_q$ if $\hat{\MM}^{(\hr)}_2=\begin{bmatrix}0&0\\\beta^{(\hr)}_3&\beta^{(\hr)}_4\end{bmatrix}$ for some $\beta^{(\hr)}_3,\beta^{(\hr)}_4\in\CF_q$.
    Since 
    \begin{equation}
    \hat{\MH}^{(\hr)}_{i}=\begin{bmatrix}\hat{\MM}^{(\hr)}_1&\hat{\MM}^{(\hr)}_2\end{bmatrix}\begin{bmatrix}\MI\\\MA_{i}\end{bmatrix}=\hat{\MM}^{(\hr)}_1+\hat{\MM}^{(\hr)}_2\MA_{i}=\MI
    \end{equation}
    and $\hat{\MM}^{(\hr)}_2=\begin{bmatrix}0&0\\\beta^{(\hr)}_3&\beta^{(\hr)}_4\end{bmatrix}$,
    $\hat{\MM}^{(\hr)}_1=\hat{\MM}^{(\hr)}_2\MA_{i}+\MI=\begin{bmatrix}1&0\\\beta^{(\hr)}_1&\beta^{(\hr)}_2\end{bmatrix}$ for some $\beta^{(\hr)}_1,\beta^{(\hr)}_2\in\CF_q$.
    
    Therefore, we can conclude that 
    $\hat{\MM}^{(\hr)}=\begin{bmatrix}1&0&0&0\\\beta^{(\hr)}_1&\beta^{(\hr)}_2&\beta^{(\hr)}_3&\beta^{(\hr)}_4\end{bmatrix}$, 
   if $\exists j_1,j_2\in[N]$ and $j_1\neq j_2$ such that $\hat{\MH}^{(\hr)}_{j_1},\hat{\MH}^{(\hr)}_{j_2}\in\CL$. 
   
   A similar proof can be made for (3).
\end{IEEEproof}

\vspace{1cm}
\begin{theorem}\label{thm:sametype}
 Given a $P_R$ and let $\CJ_i\triangleq\{j:j\in[N]\mbox{ and } B_{i,j}(P_R)=1\}$,  if $B_i(P_R)<2K$ for some $i\in\CI^{(r)}$, then either $\MH^{(r)}_j\in\CL$ for all $j\in\CJ_i$ or  $\MH^{(r)}_j\in\CR$ for all $j\in\CJ_i$.
\end{theorem}
\begin{IEEEproof}
    From Lemma \ref{lm:Bicalculation} and $B_i(P_R)<2K$, 
    we can find at least $3$ different indexes from $[N]\setminus\CI^{(r)}$, $j_1$, $j_2$, and $j_3$, such that $\MH^{(r)}_{j_1},\MH^{(r)}_{j_2},\MH^{(r)}_{j_3}\in\{\CL\cup\CR\}$. Hence, $|\CJ_i|\ge 3$. 
    By pigeonhole principle, there must be two of $\left\{\MH^{(r)}_{j_1},\MH^{(r)}_{j_2},\MH^{(r)}_{j_3}\right\}$ are either in $\CL$ or $\CR$.  
    
    WLOG, we assume $\MH^{(r)}_{j_1},\MH^{(r)}_{j_2}\in\CL$. 
    By Lemma \ref{lm:hatM}, we can construct a repair matrix
    \begin{equation}
        \hat{\MM}^{(r)}=\begin{bmatrix}1&0&0&0\\\beta^{(r)}_1&\beta^{(r)}_2&\beta^{(r)}_3&\beta^{(r)}_4\end{bmatrix}, 
    \end{equation}
    where $\hat{\MM}^{(r)}=\left[\MH^{(r)}_i\right]^{-1}\MM^{(r)}$ and $\beta^{(r)}_k\in\CF_q$ for all $k\in[4]$. 
    Suppose there is a $\hat{j}\in[N]\setminus\{j_1,j_2\}$ such that $\MH^{(r)}_{\hat{j}}\in\CR$, then
    \begin{equation}
        \MH^{(r)}_{\hat{j}}
        =\MM^{(r)}\begin{bmatrix}\MI\\\MA_{\hat{j}}\end{bmatrix}
        =\MH^{(r)}_i\hat{\MM}^{(r)}\begin{bmatrix}\MI\\\MA_{\hat{j}}\end{bmatrix}        
        =\MH^{(r)}_i\begin{bmatrix}1&0\\\beta_1&\beta_2\end{bmatrix}
        =\MH^{(r)}_i\hat{\MH}^{(r)}_{\hat{j}}
    \end{equation}
    for some $\beta_1,\beta_2\in\CF_q$. 
    Since $\hat{\MH}^{(r)}_{\hat{j}}\in\{\CL\cup\CR\}$, we have $\beta_2=0$ and $\hat{\MH}^{(r)}_{\hat{j}}\in\CL$. 
    Therefore, $\MH^{(r)}_{\hat{j}}=\MH^{(r)}_i\hat{\MH}^{(r)}_{\hat{j}}\in\CL$, which contradicts to the assumption of $\MH^{(r)}_{\hat{j}}\in\CR$. 
    Therefore, we can prove that, for all $j\in\CJ_i$, $\MH^{(r)}_j\in\CL$. 

  A similar proof can be made for $\MH^{(r)}_j\in\CR$ for all $j\in\CJ_i$  when we assume  $\MH^{(r)}_{j_1}, \MH^{(r)}_{j_2}\in\CR$.  
\end{IEEEproof}

\vspace{1cm}
\section{Lower bound on $B(P_R)$}
Let $\CP_R$ be the set of the all {\it effective}\footnote{From Theorem \ref{thm:girth4free}, we know that $P_R$ can be reduced to $P_{R-1}$ if there exists different $r_1,r_2\in[R]$ and different $i,j\in[N]$ such that $\MH^{(r_1)}_i,\MH^{(r_2)}_i\in\CL$ (or $\CR$) and $\MH^{(r_1)}_j,\MH^{(r_2)}_j\in\CL$ (or $\CR$).
For those repair processes of size $R$ satisfy the above constraint are said to be {\it not effective} and excluded from $\CP_R$ to avoid duplicate computation.} 
repair processes of size $R$, our goal is to minimizes the total repair bandwidth over $\CP_R$ for a given $R$, i.e.,
\begin{equation}
 B^*(R)\triangleq\min_{P_R\in\CP_R}B(P_R). 
\end{equation}
Let 
\begin{equation}
    \CJ_r(P_R)\triangleq\left\{j:j\in[N]\mbox{ and }\MH^{(r)}_j\in\{\CL\cup\CR\}\right\},
\end{equation}
as stated in Lemmas \ref{lm:totalrepairbandwidth} and \ref{lm:Bicalculation},
the total repair bandwidth of $P_R$ is 
\begin{IEEEeqnarray}{rCl}
    B(P_R)
    &=&\sum_{r\in[R]}|\CI^{(r)}|\times\Big[2(N-1)-|\CJ_r(P_R)|\Big]\\
    &=&\sum_{r\in[R]}|\CI^{(r)}|\times\Big[2(N-1)\Big]-\sum_{r\in[R]}|\CI^{(r)}|\times|\CJ_r(P_R)|\\
    &=&N\times\Big[2(N-1)\Big]-\sum_{r\in[R]}|\CI^{(r)}|\times|\CJ_r(P_R)|.
\end{IEEEeqnarray} 
Therefore, 
\begin{equation}\label{eq:mintomax}
    B^*(R)=\min_{P_R\in\CP_R}B(P_R)=N\times\Big[2(N-1)\Big]-\max_{P_R\in\CP_R}\sum_{r\in[R]}|\CI^{(r)}|\times|\CJ_r(P_R)|,
\end{equation}
which means that minimizing $B(P_R)$ over $P_R\in\CP_R$ is equivalent to maximizing $\sum_{r\in[R]}|\CI^{(r)}|\times|\CJ_r(P_R)|$ over $P_R\in\CP_R$.

For any repair process $P_R$, we define $\MH_{P_R}$ as
\begin{equation}
 \MH_{P_R}\triangleq\begin{bmatrix}\MM^{(1)}\\\MM^{(2)}\\\vdots\\\MM^{(R)}\end{bmatrix}\MH
=\begin{bmatrix}
\MH^{(1)}_1&\MH^{(1)}_2&\cdots &\MH^{(1)}_N\\
  \MH^{(2)}_1&\MH^{(2)}_2&\cdots &\MH^{(2)}_N\\
  \vdots&\vdots&\ddots&\vdots\\
   \MH^{(R)}_1&\MH^{(R)}_2&\cdots &\MH^{(R)}_N
 \end{bmatrix}.
 \end{equation}
 From Theorem \ref{thm:sametype}, we know that either $\MH^{(r)}_i\in\CL$ for all $i\in\CJ_r(P_R)$ or $\MH^{(r)}_i\in\CR$ for all $i\in\CJ_r(P_R)$. Hence we can categorize all repair matrices into $2$ types. 
 Since we can swap any $\MM^{(i)}$ and $\MM^{(j)}$, $i,j\in [R]$, without effecting the total repair bandwidth, 
WLOG, we assume that there is a $0\leq R_\textL\leq R$ such that 
\begin{equation}
    \MH^{(r)}_k\in\CL\mbox{ if }k\in\CJ_r(P_R)\mbox{ and }r\in[R_\textL]
\end{equation}
and 
\begin{equation}
\MH^{(r)}_k\in\CR\mbox{ if } k\in\CJ_r(P_R)\mbox{ and }r\in[R]\setminus[R_\textL].
\end{equation}
 
 Since we can also swap any $\MA_i$ and $\MA_j$ in $\MH$ without effecting the total repair bandwidth of $P_R$, 
 WLOG, we can assume that any $i\in\CI^{(r)}$ is less than any $j\in\CI^{(r+1)}$ for all $r\in[R-1]$. Let $\ell_r\triangleq|\CI^{(r)}|$ for all $r\in[R]$, then $\{\CI^{(r)}\}_{r\in[R]}$ can be uniquely decided by $\vec{\ell}\triangleq(\ell_1,\ell_2,\ldots,\ell_R)$ as
\begin{IEEEeqnarray}{rCl}
\CI^{(1)}&=&\left\{1,2,\ldots,\ell_1\right\}=[\ell^*_1];\\
\CI^{(2)}&=&\left\{\ell^*_1+1,\ell^*_1+2,\ldots,\ell^*_2\right\}=[\ell^*_2]\setminus[\ell^*_1];\\
&\vdots&\\
\CI^{(r)}&=&\left\{\ell^*_{r-1}+1,\ell^*_{r-1}+2,\ldots,\ell^*_{r}\right\}=[\ell^*_r]\setminus[\ell^*_{r-1}];\\
&\vdots&\\
\CI^{(R)}&=&\left\{\ell^*_{R-1}+1, \ell^*_{R-1}+2,\ldots,N\right\}=[N]\setminus[\ell^*_{R-1}],
\end{IEEEeqnarray}
where $\ell^*_r\triangleq\sum_{k\in[r]}\ell_k$. 

By summarizing Theorems \ref{thm:girth4free} and \ref{thm:sametype}, 
we can have the following constraints on $\MH_{P_R}$. 
\begin{itemize}
 \item \emph{Constraint \ref{thm:girth4free}-L}: 
 There exists no distinct $r_1,r_2\in[R_\textL]$ and distinct $i,j\in[N]$ such that $\MH^{(r_1)}_i,\MH^{(r_2)}_i,\MH^{(r_1)}_j,\MH^{(r_2)}_j\in\CL$;
 
  \item  \emph{Constraint \ref{thm:girth4free}-R}:
  There exists no distinct $r_1,r_2\in[R]\setminus [R_\textL]$ and distinct $i,j\in[N]$ such that $\MH^{(r_1)}_i,\MH^{(r_2)}_i,\MH^{(r_1)}_j,\MH^{(r_2)}_j\in\CR$;
  
 \item \emph{Constraint \ref{thm:sametype}-L}: 
 If $r\in[R_\textL]$, $\MH^{(r)}_j\in\CL$ for all $j\in\CJ_r(P_R)$;
 
  \item \emph{Constraint \ref{thm:sametype}-R}: 
  If $r\in[R]\setminus[R_\textL]$, $\MH^{(r)}_j\in\CR$ for all $j\in\CJ_r(P_R)$.
\end{itemize}
We then turn our attention to the block matrices which satisfy above constraints without considering the repair process $P_R$. 

Given $R$, $R_\textL$, and $\vec{\ell}=(\ell_1,\ell_2,\ldots,\ell_R)$ such that $1\leq R\leq N$, $0\leq R_\textL\leq R$, and $\sum_{r\in[R]}\ell_r=N$, we define a set of block matrices as 
\begin{equation}\label{eq:HRl}
 \CH(R, R_\textL, \vec{\ell}\,)=\left\{
 \begin{bmatrix}
{\MH}^{(1)}_1&{\MH}^{(1)}_2&\cdots &{\MH}^{(1)}_N\\
  {\MH}^{(2)}_1&{\MH}^{(2)}_2&\cdots &{\MH}^{(2)}_N\\
  \vdots&\vdots&\ddots&\vdots\\
   {\MH}^{(R)}_1&{\MH}^{(R)}_2&\cdots &{\MH}^{(R)}_N
 \end{bmatrix}:
 \, 
 \begin{matrix}
     {\MH}^{(r)}_{i} \mbox{ is invertible  if } i\in[\ell^*_r]\setminus[\ell^*_{r-1}];\\
     \mbox{The first $R_\textL$ rows satisfy Constraints \ref{thm:girth4free}-L and \ref{thm:sametype}-L};\\
     \mbox{The last $R-R_\textL$ rows satisfy Constraints \ref{thm:girth4free}-R and \ref{thm:sametype}-R}
 \end{matrix}
\right\},
\end{equation}
where $\ell^*_r=\sum_{k\in[r]}\ell_k$.
For every $\bar{\MH}=\begin{bmatrix}\bar{\MH}^{(r)}_i\end{bmatrix}_{r\in[R],i\in[N]}\in\CH(R, R_\textL, \vec{\ell}\,)$, similar with the definitions of $\CJ_r(P_R)$ and $B(P_R)$, we define the $\CJ_r(\bar{\MH})$ and $B(\bar{\MH})$ as
\begin{IEEEeqnarray}{rCl}
    \CJ_r(\bar{\MH})&\triangleq&\left\{j:j\in[N]\mbox{ and }\bar{\MH}^{(r)}_j\in\{\CL\cup\CR\}\right\},\\
B(\bar{\MH})&\triangleq& N\times\Big[2(N-1)\Big]-\sum_{r\in[R]}\ell_r\times|\CJ_r(\bar{\MH})|.\label{eq:BH}
\end{IEEEeqnarray}

It should be noted that, as discussed at the beginning of this section, we can construct a $P'_R$ for any repair process $P_R$ such that $B(P_R)=B(P'_R)$ and $\MH_{P'_R}\in \CH(R, R_\textL, \vec{\ell}\,)$; on the contrary, there is no guarantee that we can construct a $P_R$ such that $\MH_{P_R}=\bar{\MH}$ for any $\bar{\MH}\in\CH(R,R_\textL,\vec{\ell}\,)$.
Therefore, we can conclude that  
\begin{equation}\label{eq:optimalBH}
 B^*(R)=\min_{P_R\in\CP_R}B(P_R)
 \geq \min_{\bar{\MH}\in\CH(R)}B(\bar{\MH})\triangleq\bar{B}^*(R),
\end{equation}
where
\begin{equation}
    \CH(R)\triangleq\bigcup_{\vec{\ell}\in\{(\ell_1,\ldots,\ell_R): \sum_{r\in[R]}\ell_r=N\}}\,\bigcup_{0\le R_\textL\le R}\CH(R,R_\textL,\vec{\ell}\,).
\end{equation}
We then turn our attention to $\bar{B}^*(R)$ rather than $B^*(R)$.

\vspace{1cm}
\begin{theorem}\label{thm:optimalBH2}
    $\bar{B}^*(R=2)= 2N(N-1)-2\left\lfloor\frac{N^2}{2}\right\rfloor$.
\end{theorem}
\begin{IEEEproof}
Given an $\vec{\ell}=(\ell_1,\ell_2)$, we consider an $\bar{\MH}_2\in\CH(2,1,\vec{\ell}\,)$ such that
    \begin{equation}\label{eq:H2}
        \bar{\MH}_2=\begin{bmatrix}
        \MI^{(1)}_1&\cdots&\MI^{(1)}_{\ell_1}&\ML^{(1)}_{\ell_1+1}&\cdots&\ML^{(1)}_{N}\\
        \MR^{(2)}_1&\cdots&\MR^{(2)}_{\ell_1}&\MI^{(2)}_{\ell_1+1}&\cdots&\MI^{(2)}_{N}\end{bmatrix},
    \end{equation}
    where $\MI^{(r)}_i\in\CM_\inv$, $\ML^{(r)}_i\in\CL$, and $\MR^{(r)}_i\in\CR$.
    Since there must be at least $\ell_r'$ invertible matrices in the $r$th blockwise row of  any $\bar{\MH}\in\CH\left(R,R_\textL,\vec{\ell}'=(\ell'_1,\ldots,\ell'_R)\,\right)$, we have $|\CJ_r(\bar{\MH})|\leq N-\ell'_r$ for any $\bar{\MH}\in\CH(R,R_\textL,\vec{\ell}')$.  
    Since $|\CJ_{r}(\bar{\MH}_2)|=N-\ell_r$ for all $r\in[2]$, the total repair bandwidth of  any $\bar{\MH}\in\CH(2,R_\textL,\vec{\ell}\,)$ is lower bounded as
    \begin{IEEEeqnarray}{rCl}
        B(\bar{\MH})
        &=&2N(N-1)-\ell_1|\CJ_{1}(\bar{\MH})|-\ell_2|\CJ_{2}(\bar{\MH})|\\
        &\ge&2N(N-1)-\ell_1(N-\ell_1)-\ell_2(N-\ell_2)\\
        &=&\min_{\bar{\MH'}\in\CH(2,R_\textL,\vec{\ell}\,)}B(\bar{\MH'})=B(\bar{\MH_2})
    \end{IEEEeqnarray}
    for any $0\le R_\textL\le 2$.
    Therefore, 
    \begin{IEEEeqnarray}{rCl}
    \bar{B}^*(R)&=&\min_{\vec{\ell}\in\{(\ell_1,\ell_2):\,\ell_1+\ell_2=N\}}\,\min_{0\le R_\textL\le 2}\,\min_{\bar{\MH}\in\CH(2,R_\textL\vec{\ell}\,)}B(\bar{\MH})\\
    &=&\min_{\vec{\ell}\in\{(\ell_1,\ell_2):\,\ell_1+\ell_2=N\}}B(\bar{\MH}_2)\\
     &=&\min_{\vec{\ell}\in\{(\ell_1,\ell_2):\,\ell_1+\ell_2=N\}} 2N(N-1)-\ell_1(N-\ell_1)-\ell_2(N-\ell_2)\\
     &=&\min_{\vec{\ell}\in\{(\ell_1,\ell_2):\,\ell_1+\ell_2=N\}} 2N(N-1)-\ell_1(N-\ell_1)-(N-\ell_1)\ell_1\\
     &=&\min_{\vec{\ell}\in\{(\ell_1,\ell_2):\,\ell_1+\ell_2=N\}} 2N(N-1)+2\ell_1^2-2N\ell_1\\
     &=&\min_{\vec{\ell}\in\{(\ell_1,\ell_2):\,\ell_1+\ell_2=N\}} 2N(N-1)+2\left(\ell_1-\frac{N}{2}\right)^2-\frac{N^2}{2}\\
     &=& 2N(N-1)-2\left\lfloor\frac{N^2}{2}\right\rfloor.
    \end{IEEEeqnarray}
\end{IEEEproof}

\vspace{1cm}
\begin{lemma}\label{lm:B2geB3}
 $\bar{B}^*(2)\ge \bar{B}^*(3)$
\end{lemma}
\begin{IEEEproof}
 Given an
 \begin{equation}\label{eq:optimalH3}
        \bar{\MH}=
        \begin{bmatrix}
        \MI^{(1)}_1&\ML^{(1)}_{2}&\cdots&\ML^{(1)}_{\ell_1}&\MB^{(1)}_{\ell_1+1}&\cdots&\MB^{(1)}_{N-1}&\ML^{(1)}_{N}\\
        \ML^{(2)}_1&\MI^{(2)}_{2}&\cdots&\MI^{(2)}_{\ell_1}&\ML^{(2)}_{\ell_1+1}&\cdots&\ML^{(2)}_{N-1}&\ML^{(2)}_{N}\\
        \MR^{(3)}_1&\MR^{(3)}_{2}&\cdots&\MR^{(3)}_{\ell_1}&\MI^{(3)}_{\ell_1+1}&\cdots&\MI^{(3)}_{N-1}&\MI^{(3)}_{N}\\
        \end{bmatrix}\in\CH(3,2,\vec{\ell}\,),
    \end{equation}
 where $\vec{\ell}=(1,\ell_1-1,\ell_2)$, $\ell_1=\lceil\frac{N}{2}\rceil$, and $\ell_2=N-\ell_1$. We then have
 \begin{IEEEeqnarray}{rCl}
  \bar{B}^*(3)&\le& B(\bar{\MH})\\
  &=&2N(N-1)-\ell_1-(\ell_1-1)(\ell_2+1)-\ell_2(\ell_1)\\
  &=&2N(N-1)-\ell_1-(\ell_1-1)(N-\ell_1+1)-(N-\ell_1)\ell_1\\
&=&2N(N-1)-(N-\ell_1+1)(2\ell_1-1)\\
&=&\begin{cases}
    2N(N-1)-\frac{N^2}{2}-\frac{N}{2}+1&\mbox{if $N$ is even};\\
    2N(N-1)-\frac{N^2}{2}-\frac{N}{2}&\mbox{if $N$ is odd}
   \end{cases}\\
   &\le&2N(N-1)-\frac{N^2}{2}\\
   &\le&\bar{B}^*(2).
 \end{IEEEeqnarray}

\end{IEEEproof}

\vspace{1cm}
\begin{lemma}\label{lm:optimalBH3}
    Given an $\vec{\ell}=(\ell_1,\ell_2,\ell_3)$ such that $\sum_{r\in[3]}\ell_r=N$ and $\ell_1\le\ell_2$.
    Let
    \begin{equation}
        \bar{\MH}_3=
        \begin{bmatrix}
        \MI^{(1)}_1&\cdots&\MI^{(1)}_{\ell^*_1}&\ML^{(1)}_{\ell^*_1+1}&\cdots&\ML^{(1)}_{\ell^*_2}&\MB^{(1)}_{\ell^*_2+1}&\cdots&\MB^{(1)}_{N-1}&\ML^{(1)}_{N}\\
        \ML^{(2)}_1&\cdots&\ML^{(2)}_{\ell^*_1}&\MI^{(2)}_{\ell^*_1+1}&\cdots&\MI^{(2)}_{\ell^*_2}&\ML^{(2)}_{\ell^*_2+1}&\cdots&\ML^{(2)}_{N-1}&\ML^{(2)}_{N}\\
        \MR^{(3)}_1&\cdots&\MR^{(3)}_{\ell^*_1}&\MR^{(3)}_{\ell^*_1+1}&\cdots&\MR^{(3)}_{\ell^*_2}&\MI^{(3)}_{\ell^*_2+1}&\cdots&\MI^{(3)}_{N-1}&\MI^{(3)}_{N}\\
        \end{bmatrix}\in\CH(3,2,\vec{\ell}\,),
    \end{equation}
    where $\ell^*_r=\sum_{k\in[r]}\ell_k$, $\MI^{(r)}_i\in\CM_\inv$, $\MB^{(r)}_i\in\CM$, $\ML^{(r)}_i\in\CL$, and $\MR^{(r)}_i\in\CR$. 
    Then,
    \begin{equation}
        B(\bar{\MH}_3)=\min_{0\le R_\textL\le R}\,\min_{\bar{\MH}\in\CH(3,R_\textL,\vec{\ell}\,)}B(\bar{\MH}).
    \end{equation}
\end{lemma}
\begin{IEEEproof}
    According to \eqref{eq:HRl}, the constraints (\ref{thm:girth4free}-L and \ref{thm:sametype}-L) on the first $R_\textL$ rows of $\bar{\MH}$ are different from the constraints (\ref{thm:girth4free}-R and \ref{thm:sametype}-R) on the last $R-R_\textL$ rows of $\bar{\MH}$. 
    Therefore, the design of the first $R_\textL$ row is irrelevant to the design of the remaining $R-R_\textL$ rows and vice versa.
   Also, for any $\bar{\MH}\in\CH(R,R_\textL,\vec{\ell}\,)$, we can replace the matrices in $\CL$ with the matrices in $\CR$ and the matrices in $\CR$ with the matrices in $\CL$ without effecting the total repair bandwidth of $\bar{\MH}$, 
    hence we have 
    \begin{equation}\label{eq:RLeqR-RL}
     \min_{\bar{\MH}\in\CH(R,R_\textL,\vec{\ell}\,)}B(\bar{\MH})=\min_{\bar{\MH}\in\CH(R,R-R_\textL,\vec{\ell}'\,)}B(\bar{\MH}),
    \end{equation}
    where $\vec{\ell}=(\ell_1,\ell_2,\ldots,\ell_R)$ and $\vec{\ell}'=(\ell_{R_\textL+1}, \ell_{R_\textL+2},\ldots,\ell_{R},\ell_{1},\ell_{2},\ldots,\ell_{R_\textL})$.
    WLOG, we assume $R_\textL\ge R/2 $.
    
    We first prove that, for any $\bar{\MH}\in\CH(3,3,\vec{\ell}\,)$, we can find an $\bar{\MH}'\in\CH(3,2,\vec{\ell}\,)$ such that $B(\bar{\MH})\ge B(\bar{\MH}')$. 
    If $\bar{\MH}$ is with $R_\textL=3$, we can construct $\bar{\MH}'$ by replacing $\bar{\MH}^{(3)}_k$, for all $k\in[\ell^*_2]$, with matrices in $\CR$ without violating any constraint. Then
    \begin{IEEEeqnarray}{rCl}
     B(\bar{\MH})
     &=&2N(N-1)-\sum_{r\in[3]}\ell_r|\CJ_r(\bar{\MH})|\\
     &\ge&2N(N-1)-\sum_{r\in[2]}\ell_r|\CJ_r(\bar{\MH})|-\ell_3(N-\ell^*_2)=B(\bar{\MH}').
    \end{IEEEeqnarray}
    Therefore, we only need to consider those $\bar{\MH}\in\CH(3,2,\vec{\ell}\,)$ while minimizing the total repair bandwidth.
    
    Let $\bar{\MH}$ be with $R_\textL=2$, as discussed before, we can have $\bar{\MH}^{(3)}_k\in\CR$ for all $k\in[\ell^*_2]$ to minimize $|\CJ_3(\bar{\MH})|$ to be $N-\ell_3$. 
    We then consider the first two rows of $\bar{\MH}$.
    All those $\bar{\MH}^{(1)}_k$ for all $k\in[\ell^*_2]\setminus[\ell^*_1]$ and  
    $\bar{\MH}^{(2)}_k$ for all $k\in[\ell^*_1]$ can be in $\CL$ without violating Constraint \ref{thm:girth4free}-L. Hence, we have 
    \begin{equation}\label{eq:H3example}
        \bar{\MH}=
        \begin{bmatrix}
        \MI^{(1)}_1&\cdots&\MI^{(1)}_{\ell^*_1}&\ML^{(1)}_{\ell^*_1+1}&\cdots&\ML^{(1)}_{\ell^*_2}&\bar{\MH}^{(1)}_{\ell^*_2+1}&\cdots&\bar{\MH}^{(1)}_{N}\\
        \ML^{(2)}_1&\cdots&\ML^{(2)}_{\ell^*_1}&\MI^{(2)}_{\ell^*_1+1}&\cdots&\MI^{(2)}_{\ell^*_2}&\bar{\MH}^{(2)}_{\ell^*_2+1}&\cdots&\bar{\MH}^{(2)}_{N}\\
        \MR^{(3)}_1&\cdots&\MR^{(3)}_{\ell^*_1}&\MR^{(3)}_{\ell^*_1+1}&\cdots&\MR^{(3)}_{\ell^*_2}&\MI^{(3)}_{\ell^*_2+1}&\cdots&\MI^{(3)}_{N}
        \end{bmatrix},
    \end{equation}
    where its submatrix 
    \begin{equation}
    \begin{bmatrix}\label{eq:H3submatrixexample}
        \bar{\MH}^{(1)}_{\ell^*_2+1}&\cdots&\bar{\MH}^{(1)}_{N}\\
        \bar{\MH}^{(2)}_{\ell^*_2+1}&\cdots&\bar{\MH}^{(2)}_{N}
    \end{bmatrix}
    \end{equation}
    must satisfy Constraint \ref{thm:girth4free}-L. 
    Let $J_r\triangleq|\{k: k\in[N]\setminus[\ell^*_2]\mbox{ and }\bar{\MH}^{(r)}_k\in\CL\}|$, if $J_1+J_2\ge (N-\ell^*_2)+2=\ell_3+2$, then the submatrix in \eqref{eq:H3submatrixexample} must violate Constraint \ref{thm:girth4free}-L according to the pigeonhole principle, which induces $J_1+J_2\leq \ell_3+1$.
    Therefore, 
    \begin{IEEEeqnarray}{rCl}
     B(\bar{\MH})
     &=&2N(N-1)-\ell_1(\ell_2+J_1)-\ell_2(\ell_1+J_2)-\ell_3(N-\ell_3)\\
     &\ge&2N(N-1)-\ell_1\left[\ell_2+(\ell_3+1-J_2)\right]-\ell_2(\ell_1+J_2)-\ell_3(N-\ell_3)\\
     &=&2N(N-1)-2\ell_1\ell_2-\ell_1\ell_3-\ell_1-\ell_3(N-\ell_3)+(\ell_1-\ell_2)J_2,\label{eq:H3J2}
    \end{IEEEeqnarray}
    where \eqref{eq:H3J2} can be minimized by maximizing $J_2$ when $\ell_1\le\ell_2$.
    Since $J_2\le \ell_3$ by \eqref{eq:H3submatrixexample} and $J_1+J_2\le\ell_3+1$, we can minimize $\eqref{eq:H3J2}$ by assigning $J_2=\ell_3$ and $J_1=1$ as $\bar{\MH}_3$ did in \eqref{eq:optimalH3}.  
\end{IEEEproof}

\vspace{1cm}
\begin{theorem}\label{thm:optimalBH3}
\begin{equation}\label{eq:optimalB3}
 \bar{B}^*(3)\geq \min_{\vec{\ell}\in\{(\ell_1,\ell_2,\ell_3):\, \ell_1+\ell_2+\ell_3=N\mbox{ and }\ell_1\le\ell_2 \}}2N(N-1)-N^2+\ell_1^2+\ell^2_2+\ell_3^2+\ell_1(\ell_3-1).
 \end{equation}
\end{theorem}
\begin{IEEEproof}
We have learned from Lemma \ref{lm:optimalBH3} that 
\begin{equation}
\min_{0\le R_\textL\le 3}\,\min_{\bar{\MH}\in\CH\left(3,R_\textL,\vec{\ell}=(\ell_1,\ell_2,\ell_3)\right)}B(\bar{\MH})
\end{equation}
can be achieved by the $\bar{\MH}_3$ in \eqref{eq:optimalH3} while $\ell_1\le\ell_2$. 
For those $\bar{\MH}'\in\CH(3,R_\textL,\vec{\ell}')$, 
where $\vec{\ell}'=(\ell'_1,\ell'_2,\ell'_3)=(\ell_2,\ell_1,\ell_3)$, we can achieve 
\begin{equation}
\min_{0\le R_\textL\le 3}\,\min_{\bar{\MH}\in\CH\left(3,R_\textL,\vec{\ell}'=(\ell_2,\ell_1,\ell_3)\right)}B(\bar{\MH})
\end{equation}
by 
\begin{equation}
        \bar{\MH}'_3=
        \begin{bmatrix}
        \MI^{(1)}_1&\cdots&\MI^{(1)}_{\ell'^*_1}&\ML^{(1)}_{\ell'^*_1+1}&\cdots&\ML^{(1)}_{\ell'^*_2}&\ML^{(1)}_{\ell'^*_2+1}&\cdots&\ML^{(1)}_{N-1}&\ML^{(1)}_{N}\\
        \ML^{(2)}_1&\cdots&\ML^{(2)}_{\ell'^*_1}&\MI^{(2)}_{\ell'^*_1+1}&\cdots&\MI^{(2)}_{\ell'^*_2}&\MB^{(2)}_{\ell'^*_2+1}&\cdots&\MB^{(2)}_{N-1}&\ML^{(2)}_{N}\\
        \MR^{(3)}_1&\cdots&\MR^{(3)}_{\ell'^*_1}&\MR^{(3)}_{\ell'^*_1+1}&\cdots&\MR^{(3)}_{\ell'^*_2}&\MI^{(3)}_{\ell'^*_2+1}&\cdots&\MI^{(3)}_{N-1}&\MI^{(3)}_{N}\\
        \end{bmatrix}
    \end{equation}
    and $B(\bar{\MH}'_3)=B(\bar{\MH}_3)$. 

Therefore, we conclude that 
\begin{IEEEeqnarray}{rCl}
\bar{B}^*(3)
&=&\min_{\vec{\ell}\in\{(\ell_1,\ell_2,\ell_3):\, \ell_1+\ell_2+\ell_3=N\}}\,\min_{0\le R_\textL\le R}\,\min_{\bar{\MH}\in\CH(3,R_\textL,\vec{\ell}\,)}B(\bar{\MH})\\
&=&\min_{\vec{\ell}\in\{(\ell_1,\ell_2,\ell_3):\, \ell_1+\ell_2+\ell_3=N\mbox{ and }\ell_1\le\ell_2\}}B(\bar{\MH}_3)\\
&=&\min_{\vec{\ell}\in\{(\ell_1,\ell_2,\ell_3):\, \ell_1+\ell_2+\ell_3=N\mbox{ and }\ell_1\le\ell_2\}} 2N(N-1)-\ell_1(\ell_2+1)-\ell_2(N-\ell_2)-\ell_3(N-\ell_3)\\
&=&\min_{\vec{\ell}\in\{(\ell_1,\ell_2,\ell_3):\, \ell_1+\ell_2+\ell_3=N\mbox{ and }\ell_1\le\ell_2\}} 2N(N-1)\nonumber\\
&&\hfill-\ell_1(N-\ell_1-\ell_3+1)-\ell_2(N-\ell_2)-\ell_3(N-\ell_3)\\
&=&\min_{\vec{\ell}\in\{(\ell_1,\ell_2,\ell_3):\, \ell_1+\ell_2+\ell_3=N\mbox{ and }\ell_1\le\ell_2\}} 2N(N-1)\nonumber\\
&&\hfill-(\ell_1+\ell_2+\ell_3)N+\ell_1^2+\ell_2^2+\ell_3^2+\ell_1(\ell_3-1)\\
&=&\min_{\vec{\ell}\in\{(\ell_1,\ell_2,\ell_3):\, \ell_1+\ell_2+\ell_3=N\mbox{ and }\ell_1\le\ell_2\}} 2N(N-1)-N^2+\ell_1^2+\ell_2^2+\ell_3^2+\ell_1(\ell_3-1).
\end{IEEEeqnarray}
\end{IEEEproof}

\vspace{1cm}
We then look into $\bar{B}^*(R)$ when $R>3$.
From Constraints \ref{thm:girth4free}-L, \ref{thm:girth4free}-R, \ref{thm:sametype}-L, and \ref{thm:sametype}-R
we notice that, among all $\bar{\MH}\in\CH(R,R_\textL,\vec{\ell}\,)$, we can minimize the total bandwidth by separately minimizing bandwidth of the first $R_\textL$ rows in $\bar{\MH}$ and the bandwidth of the remaining $R-R_\textL$ rows  in $\bar{\MH}$, i.e.,
\begin{IEEEeqnarray}{rCl}
 &&\min_{\bar{\MH}\in\CH\left(R,R_\textL,\vec{\ell}=(\ell_1,\ldots,\ell_R)\right)}B(\bar{\MH})\\
 &=&\min_{\bar{\MH}\in\CH\left(R,R_\textL,\vec{\ell}=(\ell_1,\ldots,\ell_R)\right)}2N(N-1)-\sum_{r\in[R]}\ell_r|\CJ_r(\bar{\MH})|\\
 &=&\min_{\bar{\MH}\in\CH\left(R,R_\textL,\vec{\ell}=(\ell_1,\ldots,\ell_R)\right)}2R_\textL(N-1)-\sum_{r\in[R_\textL]}\ell_r|\CJ_r(\bar{\MH})|\\
 &&+\min_{\bar{\MH}\in\CH\left(R,R_\textL,\vec{\ell}=(\ell_1,\ldots,\ell_R)\right)}2(R-R_\textL)(N-1)-\sum_{r\in[R]\setminus [R_\textL]}\ell_r|\CJ_r(\bar{\MH})|.
\end{IEEEeqnarray}
Also as discussed in \eqref{eq:mintomax}, we can turn the minimization problem in to a maximization program. Therefore, we first focus on the following problem,
\begin{equation}\label{eq:maxRL}
\max_{\bar{\MH}\in\CH\left(R,R_\textL,\vec{\ell}=(\ell_1,\ldots,\ell_R)\right)}\sum_{r\in[R_\textL]}\ell_r|\CJ_r(\bar{\MH})|.
\end{equation}
The following lemma further simplifies \eqref{eq:maxRL}.

\begin{lemma}
\begin{equation}
 \max_{\bar{\MH}\in\CH\left(R,R_\textL,\vec{\ell}=(\ell_1,\ldots,\ell_R)\right)}\sum_{r\in[R_\textL]}\ell_r|\CJ_r(\bar{\MH})|=
 \max_{\bar{\MH}\in\CH\left(R_\textL+1,R_\textL,\vec{\ell}_\textL=(\ell_1,\ldots,\ell_{R_\textL},N-\ell^*_{R_\textL})\right)}\sum_{r\in[R_\textL]}\ell_r|\CJ_r(\bar{\MH})|,
\end{equation}
where $\ell^*_r=\sum_{i\in[r]}\ell_i$.
 \end{lemma}
\begin{IEEEproof}
 Since the summation of $\sum_{r\in[R_\textL]}\ell_r|\CJ_r(\bar{\MH})|$ only considers the first $R_\textL$ rows, the maximization is relevant to $\sum_{r\in[R]\setminus[R_\textL]}\ell_r$ but irrelevant of those individual $\ell_r$ for $r\in[R]\setminus[R_\textL]$.  
\end{IEEEproof}

\vspace{1cm}
\begin{lemma}\label{lm:DL2}
Given $R_\textL\ge 3$ and $\vec{\ell}=\{\ell_1,\ell_2,\ldots,\ell_{R_\textL}, N-\ell^*_{R_\textL}\}$, for any $\bar{\MH}\in\CH\left(R_\textL+1,R_\textL,\vec{\ell}\,\right)$, we can always construct a $\bar{\MH}'\in\CH\left(3,2,\vec{\ell}'=(\ell'_1, \ell'_2,  N-\ell^*_{R_\textL})\right)$ such that
\begin{equation}
 \sum_{r\in[R_\textL]}\ell_r|\CJ_r(\bar{\MH})|\le\sum_{r\in[2]}\ell'_r|\CJ_r(\bar{\MH}')|.
\end{equation}
\end{lemma}
\begin{IEEEproof}
 To simplify the problem formulation, for any $\bar{\MH}\in\CH\left(R_\textL+1,R_\textL,\vec{\ell}=(\ell_1,\ldots,\ell_{R_\textL+1})\right)$, we define  $J^{(r)}_i(\bar{\MH})$ as 
 \begin{equation}
  J^{(r)}_i(\bar{\MH})\triangleq\left|\left\{j: j\in[\ell^*_{i}]\setminus[\ell^*_{i-1}]\mbox{ and }\bar{\MH}^{(r)}_j\in\CL\right\}\right|,
 \end{equation}
 which denotes the number of block matrices in $\begin{bmatrix}\bar{\MH}^{(r)}_{\ell^*_{i-1}+1}&\bar{\MH}^{(r)}_{\ell^*_{i-1}+2}&\cdots&\bar{\MH}^{(r)}_{\ell^*_{i}}\end{bmatrix}$ belongs to $\CL$. 
 By definition, we have $J^{(r)}_r(\bar{\MH})=0$ and $|\CJ_r(\bar{\MH})|=\sum_{i\in[R+1]}J^{(r)}_i(\bar{\MH})$.

 \begin{table}[ht]\centering
 \caption{ $J^{(r)}_i(\bar{\MH})$ for $\bar{\MH}\in\CH\left(4,3,\vec{\ell}=(\ell_1,\ell_2,\ell_{3},N-\ell^*_{3})\right)$. }
 \label{tbl:HtableDL3}
  \begin{tabular}{|c|c|c|c|}\hline
   $\ell_1$&$\ell_2$&$\ell_3$&$N-\ell^*_{3}$\\\hline\hline
  \color{red}{$J^{(1)}_1(\bar{\MH})=0$} & $J^{(1)}_2(\bar{\MH})$ & $J^{(1)}_3(\bar{\MH})$ & $J^{(1)}_4(\bar{\MH})$\\\hline
   $J^{(2)}_1(\bar{\MH})$ &\color{red}{$J^{(2)}_2(\bar{\MH})=0$} & $J^{(2)}_3(\bar{\MH})$ & $J^{(2)}_4(\bar{\MH})$\\\hline
   $J^{(3)}_1(\bar{\MH})$ & $J^{(3)}_2(\bar{\MH})$ & \color{red}{$J^{(3)}_3(\bar{\MH})=0$} & $J^{(3)}_4(\bar{\MH})$\\\hline
  \end{tabular}
 \end{table}
 
 Let us consider the case of $R_\textL=3$, i.e., $\bar{\MH}\in\CH\left(4,3,\vec{\ell}=(\ell_1,\ell_2,\ell_3,N-\ell^*_{3})\right)$.
 WLOG, we asumme $\ell_1\le\ell_2\le\ell_3$.
 The corresponding $J^{(r)}_i(\bar{\MH})$ for all $r\in[3]$ and $i\in[4]$ are listed in Table \ref{tbl:HtableDL3}, and we have
 \begin{multline}\label{eq:DL3}
  \sum_{r\in[3]}\ell_r|\CJ_r(\bar{\MH})|=
  \ell_1\left[J^{(1)}_2(\bar{\MH})+J^{(1)}_3(\bar{\MH})+J^{(1)}_4(\bar{\MH})\right]\\+
  \ell_2\left[J^{(2)}_1(\bar{\MH})+J^{(2)}_3(\bar{\MH})+J^{(2)}_4(\bar{\MH})\right]+
  \ell_3\left[J^{(3)}_1(\bar{\MH})+J^{(3)}_2(\bar{\MH})+J^{(3)}_4(\bar{\MH})\right].
 \end{multline}
 Combining Constraint \ref{thm:girth4free}-L and pigeonhole principle, we can have the following inequalities from Table \ref{tbl:HtableDL3},
\begin{IEEEeqnarray}{rCl}\label{eq:ineq-1}
 J^{(1)}_2(\bar{\MH})+ J^{(1)}_4(\bar{\MH})+ J^{(3)}_2(\bar{\MH})+ J^{(3)}_4(\bar{\MH})&\le&\ell_2+(N-\ell^*_3)+1\\\label{eq:ineq-2}
 J^{(2)}_1(\bar{\MH})+ J^{(2)}_4(\bar{\MH})+ J^{(3)}_1(\bar{\MH})+ J^{(3)}_4(\bar{\MH})&\le&\ell_1+(N-\ell^*_3)+1\\\label{eq:ineq-3}
 J^{(1)}_3(\bar{\MH})+ J^{(1)}_4(\bar{\MH})+ J^{(2)}_3(\bar{\MH})+ J^{(2)}_4(\bar{\MH})&\le&\ell_3+(N-\ell^*_3)+1 \\\label{eq:ineq-4}
 J^{(1)}_3(\bar{\MH})+ J^{(2)}_3(\bar{\MH})&\le&\ell_3+1\\\label{eq:ineq-5}
 J^{(1)}_2(\bar{\MH})+J^{(3)}_2(\bar{\MH})&\le&\ell_2+1\\\label{eq:ineq-6}
 J^{(2)}_1(\bar{\MH})+J^{(3)}_1(\bar{\MH})&\le&\ell_1+1.
\end{IEEEeqnarray}
Then we can have an upper bound of \eqref{eq:DL3} by applying \eqref{eq:ineq-2}, \eqref{eq:ineq-4}, and \eqref{eq:ineq-5} as
\begin{IEEEeqnarray}{rCl}
 &&\sum_{r\in[3]}\ell_r|\CJ_r(\bar{\MH})|\nonumber\\
 &\le&\ell_1\left[(\ell_2+1)-J^{(3)}_2(\bar{\MH})+(\ell_3+1)-J^{(2)}_3(\bar{\MH})+J^{(1)}_4(\bar{\MH})\right]\nonumber\\
 &&+\ell_2\left[\Big(\ell_1+N-\ell^*_3+1\Big)-J^{(3)}_1(\bar{\MH})-J^{(3)}_4(\bar{\MH})+J^{(2)}_3(\bar{\MH})\right]\nonumber\\\label{eq:DL3UB-1}
 &&+\ell_3\left[J^{(3)}_1(\bar{\MH})+J^{(3)}_2(\bar{\MH})+J^{(3)}_4(\bar{\MH})\right]\\\label{eq:DL3UB-2}
 &=&\ell_1\left[\ell_2+\ell_3+2+J^{(1)}_4(\bar{\MH})\right]+
 \ell_2\Big[\ell_1+N-\ell^*_3+1\Big]\nonumber\\
 &&+(\ell_3-\ell_1)J^{(3)}_2(\bar{\MH})+(\ell_2-\ell_1)J^{(2)}_3(\bar{\MH})+(\ell_3-\ell_2)J^{(3)}_1(\bar{\MH})+(\ell_3-\ell_2)J^{(3)}_4(\bar{\MH})
 \\\label{eq:DL3UB-3}
 &\le&\ell_1\left[\ell_2+\ell_3+2+J^{(1)}_4(\bar{\MH})\right]+
 \ell_2\Big[\ell_1+N-\ell^*_3+1\Big]\nonumber\\
 &&+(\ell_3-\ell_1)\ell_2+(\ell_2-\ell_1)\ell_3+(\ell_3-\ell_2)\ell_1+(\ell_3-\ell_2)(N-\ell^*_3)\\\label{eq:DL3UB-4}
 &=&\ell_1\left[2+J^{(1)}_4(\bar{\MH})\right]+
 \ell_2\Big[\ell_3+1\Big]+
 \ell_3\Big[\ell_1+\ell_2+N-\ell^*_3\Big],
\end{IEEEeqnarray}
where \eqref{eq:DL3UB-3} is due to $\ell_1\le\ell_2\le\ell_3$, $J^{(3)}_2(\bar{\MH})\le\ell_2$, $J^{(2)}_3(\bar{\MH})\le\ell_3$, $J^{(3)}_1(\bar{\MH})\le\ell_1$, and $J^{(3)}_4(\bar{\MH})\le N-\ell^*_3$.
We then investigate the $\bar{\MH}$ achieves \eqref{eq:DL3UB-4} as Table \ref{tbl:HtableDL3-1} shown. 

 \begin{table}[ht]\centering
 \caption{ $J^{(r)}_i(\bar{\MH})$ for the $\bar{\MH}$ achieves \eqref{eq:DL3UB-4}. }
 \label{tbl:HtableDL3-1}
  \begin{tabular}{|c|c|c|c|}\hline
   $\ell_1$&$\ell_2$&$\ell_3$&$N-\ell^*_{3}$\\\hline\hline
  \color{red}{$J^{(1)}_1(\bar{\MH})=0$} & \color{red}{$J^{(1)}_2(\bar{\MH})=1$} & \color{red}{$J^{(1)}_3(\bar{\MH})=1$} & $J^{(1)}_4(\bar{\MH})$\\\hline
  \color{red}{$J^{(2)}_1(\bar{\MH})=1$} &\color{red}{$J^{(2)}_2(\bar{\MH})=0$} & \color{red}{$J^{(2)}_3(\bar{\MH})=\ell_3$} & \color{red}{$J^{(2)}_4(\bar{\MH})=0$}\\\hline
   \color{red}{$J^{(3)}_1(\bar{\MH})=\ell_1$} & \color{red}{$J^{(3)}_2(\bar{\MH})=\ell_2$} & \color{red}{$J^{(3)}_3(\bar{\MH})=0$} & \color{red}{$J^{(3)}_4(\bar{\MH})=N-\ell^*_3$}\\\hline
  \end{tabular}
 \end{table}

Due to inequality \eqref{eq:ineq-1}, we have $J^{(1)}_4(\bar{\MH})=0$. Therefore, we upper bound \eqref{eq:DL3} as
\begin{equation}\label{eq:DL3UB-final}
 \sum_{r\in[3]}\ell_r|\CJ_r(\bar{\MH})|
 \le  2\ell_1+\ell_2(\ell_3+1)+\ell_3\Big(\ell_1+\ell_2+(N-\ell^*_3)\Big).
\end{equation}

\begin{table}[ht]\centering
 \caption{$J^{(r)}_i(\bar{\MH}')$ for $\bar{\MH}'\in\CH\left(3,2,\vec{\ell}=(\ell_1+\ell_2,\ell_{3},N-\ell^*_{3})\right)$. }
 \label{tbl:HtableDL2}
  \begin{tabular}{|c|c|c|}\hline
   $\ell_1+\ell_2$&$\ell_3$&$N-\ell^*_{3}$\\\hline\hline
  \color{red}{$J^{(1)}_1(\bar{\MH}')=0$} & $J^{(1)}_2(\bar{\MH}')=\ell_3$ & $J^{(1)}_3(\bar{\MH}')=1$ \\\hline
   $J^{(2)}_1(\bar{\MH}')=\ell_1+\ell_2$ &\color{red}{$J^{(2)}_2(\bar{\MH}')=0$} & $J^{(2)}_3(\bar{\MH}')=N-\ell^*_3$\\\hline
  \end{tabular}
 \end{table}
 
Next, we construct a $\bar{\MH}'\in\CH\big(3,2,\vec{\ell}'=(\ell_1+\ell_2,\ell_3,N-\ell^*_3)\big)$ with the corresponding $J^{(r)}_i(\bar{\MH}')$ as shown in Table \ref{tbl:HtableDL2}.
Therefore
\begin{equation}\label{eq:DL2UB-final}
 \sum_{r\in[2]}\ell_r|\CJ_r(\bar{\MH}')|
 =  \ell_1(\ell_3+1)+\ell_2(\ell_3+1)+\ell_3\Big(\ell_1+\ell_2+(N-\ell^*_3)\Big), 
\end{equation}
By subtracting \eqref{eq:DL3UB-final} by \eqref{eq:DL2UB-final}, we have
\begin{equation}
    \sum_{r\in[3]}\ell_r|\CJ_r(\bar{\MH})|-\sum_{r\in[2]}\ell_r|\CJ_r(\bar{\MH}')|\le \ell_1(1-\ell_3)\le 0.
\end{equation}

Therefore, for any $\bar{\MH}\in\CH\left(4,3,\vec{\ell}=(\ell_1,\ell_2,\ell_3,N-\ell^*_3)\right)$, 
we can always construct an 
$\bar{\MH}'\in\CH\left(3,2,\vec{\ell}'\right)$ such that
$\vec{\ell}'_\textL=(\ell_1+\ell_2,\ell_3,N-\ell^*_3)$ and 
\begin{equation}\label{eq:RL3leRL2}
\sum_{r\in[3]}\ell_r|\CJ_r(\bar{\MH})|\le\sum_{r\in[2]}\ell_r|\CJ_r(\bar{\MH}')|.
\end{equation}

Next we consider an $\bar{\MH}\in\CH\left(R_\textL+1,R_\textL,\vec{\ell}=(\ell_1,\ell_2,\ldots,\ell_{R_\textL},N-\ell^*_{R_\textL})\,\right)$ for $R_\textL>3$, which is with the corresponding $J^{(r)}_i(\bar{\MH})$ in Table \ref{tbl:HtableRLplus2}.
 \begin{table}[ht]\centering\caption{ $J^{(r)}_i(\bar{\MH})$ for $\bar{\MH}\in\CH\left(R_\textL+1,R_\textL,\vec{\ell}=(\ell_1,\ell_2,\ldots,\ell_{R_\textL},N-\ell^*_{R_\textL})\,\right)$. }
 \label{tbl:HtableRLplus2}
  \begin{tabular}{|c|c|c|c|c|c|}\hline
   $\ell_1$&$\ell_2$&$\cdots$&$\ell_{R_\textL-1}$&$\ell_{R_\textL}$&$N-\ell^*_{R_\textL}$\\\hline\hline
  \color{red}{$J^{(1)}_1(\bar{\MH})=0$} & $J^{(1)}_2(\bar{\MH})$ &$\cdots$& $J^{(1)}_{R_\textL-1}(\bar{\MH})$ & $J^{(1)}_{R_\textL}(\bar{\MH})$& $J^{(1)}_{R_\textL+1}(\bar{\MH})$\\\hline
  $J^{(2)}_1(\bar{\MH})$ & \color{red}{$J^{(2)}_2(\bar{\MH})=0$} &$\cdots$& $J^{(2)}_{R_\textL-1}(\bar{\MH})$ & $J^{(2)}_{R_\textL}(\bar{\MH})$& $J^{(2)}_{R_\textL+1}(\bar{\MH})$\\\hline
 $\vdots$&$\vdots$&$\ddots$&$\vdots$&$\vdots$&$\vdots$\\\hline
  $J^{(R_\textL-1)}_1(\bar{\MH})$ & $J^{(R_\textL-1)}_2(\bar{\MH})$ &$\cdots$& \color{red}{$J^{(R_\textL-1)}_{R_\textL-1}(\bar{\MH})=0$} & $J^{(R_\textL-1)}_{R_\textL}(\bar{\MH})$& $J^{(R_\textL-1)}_{R_\textL+1}(\bar{\MH})$\\\hline
   $J^{(R_\textL)}_1(\bar{\MH})$ & $J^{(R_\textL)}_2(\bar{\MH})$ &$\cdots$& $J^{(R_\textL)}_{R_\textL-1}(\bar{\MH})$ & \color{red}{$J^{(R_\textL)}_{R_\textL}(\bar{\MH})=0$}& $J^{(R_\textL)}_{R_\textL+1}(\bar{\MH})$\\\hline
  \end{tabular}
 \end{table}
 As discussed in \eqref{eq:DL3UB-final}, we have 
 \begin{equation}
 \sum_{r\in\{r',R_\textL-1,R_{\textL}\}} \ell_r |\CJ_r(\bar{\MH})|\le 2\ell_{r'}+\ell_{R_\textL-1}(\ell_{R_\textL}+1)+\ell_{R_\textL}(N-\ell_{R_\textL}),
 \end{equation}
 for all $r'\in[R_\textL-2]$. 
 Hence, 
 \begin{equation}\label{eq:RLplus2}
  \sum_{r\in[R_\textL]}\ell_r |\CJ_r(\bar{\MH})|\le 2\left(\sum_{r\in[R_\textL-2]}\ell_r\right)+\ell_{R_\textL-1}(\ell_{R_\textL}+1)+\ell_{R_\textL}(N-\ell_{R_\textL})=\sum_{r\in[3]}\ell_r |\CJ_r(\bar{\MH}')|,
 \end{equation}
 where $\bar{\MH}'$ is with the corresponding $J^{(r)}_i(\bar{\MH}')$ in Table \ref{tbl:HtableDL3-2}.

\begin{table}[ht]\centering
 \caption{ $J^{(r)}_i(\bar{\MH}')$ for the $\bar{\MH}'$ in \eqref{eq:RLplus2}. }
 \label{tbl:HtableDL3-2}
  \begin{tabular}{|c|c|c|c|}\hline
   $\sum_{r\in[R_\textL-2]}\ell_r$&$\ell_{R_\textL-1}$&$\ell_{R_\textL}$&$N-\ell^*_{R_\textL}$\\\hline\hline
  $J^{(1)}_1(\bar{\MH})=0$ & $J^{(1)}_2(\bar{\MH})=1$ & $J^{(1)}_3(\bar{\MH})=1$ & $J^{(1)}_4(\bar{\MH})=0$\\\hline
  $J^{(2)}_1(\bar{\MH})=1$ &$J^{(2)}_2(\bar{\MH})=0$ & $J^{(2)}_3(\bar{\MH})=\ell_{R_\textL}$ & $J^{(2)}_4(\bar{\MH})=0$\\\hline
   $J^{(3)}_1(\bar{\MH})=\sum_{r\in[R_\textL-2]}\ell_r$ & $J^{(3)}_2(\bar{\MH})=\ell_{R_\textL-1}$ & $J^{(3)}_3(\bar{\MH})=0$ & $J^{(3)}_4(\bar{\MH})=N-\ell^*_{R_\textL}$\\\hline
  \end{tabular}
 \end{table}
 Since 
 \begin{equation}
  \bar{\MH}'\in\CH\left(4,3,\bigg(\sum_{r\in[R_\textL-2]}\ell_r,\ell_{R_\textL-1},\ell_{R_\textL},N-\ell^*_{R_\textL}\bigg)\right),
 \end{equation}
as proven in \eqref{eq:RL3leRL2}, we can find an $\bar{\MH}''\in\CH\Big(3,2,(\ell''_1,\ell''_2,N-\ell^*_{R_\textL})\Big)$ such that 
\begin{equation}
 \sum_{r\in[R_\textL+2]}\ell_r |\CJ_r(\bar{\MH})|\le
 \sum_{r\in[3]}\ell_r |\CJ_r(\bar{\MH}')|\le
 \sum_{r\in[2]}\ell_r |\CJ_r(\bar{\MH}'')|.
\end{equation}
\end{IEEEproof}

\vspace{1cm}
Next, for those remaining $R-R_\textL$ rows of $\bar{\MH}$, we can have a similar lemma below.
\begin{lemma}\label{lm:DR2}
Given $R-1\ge 3$ and $\vec{\ell}=\{\ell^*_{R-1},\ell_1,\ell_2,\ldots,\ell_{R-1}\}$, for any $\bar{\MH}\in\CH\left(R,R-1,\vec{\ell}\,\right)$, we can always construct a $\bar{\MH}'\in\CH\left(3,2,\vec{\ell}'=(\ell^*_{R-1}, \ell'_1,  \ell'_2\right)$ such that
\begin{equation}
 \sum_{r\in[R]\setminus[1]}\ell_r|\CJ_r(\bar{\MH})|\le\sum_{r\in\{2,3\}}\ell'_r|\CJ_r(\bar{\MH}')|.
\end{equation}
\end{lemma}
\begin{IEEEproof}
The proof is similar to the proof of Lemma \ref{lm:DL2}, hence we omit it here.
\end{IEEEproof}

\vspace{1cm}
Combining Theorems \ref{thm:optimalBH2}, \ref{thm:optimalBH3} and Lemmas \ref{lm:DL2}, \ref{lm:DR2}, we can lower bound $\bar{B}^*(R)$ for all $R\in[N]$ in the following theorem.
\begin{theorem}
Let 
\begin{multline}\label{eq:delta3}
 \Delta_3\triangleq \min_{\vec{\ell}\in\{(\ell_1,\ell_2,\ell_3):\, \sum_{r\in[3]}\ell_r=N\mbox{ and }\ell_1\le\ell_2 \}}2N(N-1)-N^2+\ell_1^2+\ell^2_2+\ell_3^2+\ell_1(\ell_3-1)
 \end{multline}
 and
\begin{multline}\label{eq:delta4}
 \Delta_4\triangleq \min_{\vec{\ell}\in\{(\ell_1,\ell_2,\ell_3,\ell_4):\, \sum_{r\in[4]}\ell_r=N\mbox{, }\ell_1\le\ell_2\mbox{, and }\ell_3\le\ell_4 \}}2N(N-1)-\\\ell_1(\ell_2+1)-\ell_2(N-\ell_2)-\ell_3(\ell_4+1)-\ell_4(N-\ell_4),
 \end{multline}
 then $\bar{B}^*(R)\ge\min\{\Delta_3,\Delta_4\}$ for all $R\in[N]$. 
 \end{theorem}
 \begin{IEEEproof}
 As discussed in \eqref{eq:RLeqR-RL}, we only need to consider the case of $R_\textL\ge R-R_\textL\ge 1$. 
 
 Lemma \ref{lm:B2geB3} and Theorem \ref{thm:optimalBH3} have proven $\bar{B}^*(1)\ge\bar{B}^*(2)\ge\bar{B}^*(3)\ge\Delta_3$ for the case of $R=2$. We next consider the case of $R\ge 3$. 
 
 For the case of $R\ge 3$ and $R-R_\textL=1$, any $\bar{\MH}\in\CH\left\{R,R-1,(\ell_1,\ell_2,\ldots,\ell_R)\right\}$ can be replaced by some $\bar{\MH}'\in\CH\left\{3,2,(\ell'_1,\ell'_2,\ell_R)\right\}$ such that $B(\MH)\ge B(\MH')$ due to Lemma
 \ref{lm:DL2}.  
 Therefore, $\bar{B}^*(R)\ge\bar{B}^*(3)\ge\Delta_3$ when $R\ge 3$ and $R-R_\textL=1$.
 
 The remaining case is of $R\ge 3$ and $R-R_\textL\ge 2$.
 Since $R_\textL\ge R- R_\textL$, this case is equivalent to the case of $R_\textL\ge R- R_\textL\ge 2$. 
 For the case of $R_\textL\ge R- R_\textL\ge 2$,
any $\bar{\MH}\in\CH\left\{R,R_\textL,(\ell_1,\ell_2,\ldots,\ell_R)\right\}$ can be replaced by some $\bar{\MH}'\in\CH\left\{4,2,(\ell'_1,\ell'_2,\ell'_3,\ell'_4)\right\}$ such that
 $\ell'_1+\ell'_2=\sum_{r\in[R_\textL]}\ell_r$, $\ell'_3+\ell'_4=\sum_{r\in[R]\setminus [R_\textL]}\ell_r$, $B(\MH)\ge B(\MH')$. Therefore, we can conclude that
 \begin{equation}\label{eq:BRLB}
  \bar{B}^*(R)\ge\min_{\vec{\ell}\in\{(\ell_1,\ell_2,\ell_3,\ell_4):\sum_{r\in[4]}\ell_r=N\}}\min_{\bar{\MH}\in\CH(4,2,\vec{\ell}\,)}B(\bar{\MH}),
 \end{equation}
 when $R_\textL\ge R-R_\textL\ge 2$. 
 We then prove that \eqref{eq:BRLB} can be further lower bounded by $\Delta_4$. 
 
 As discussed in the beginning of this section, we swap rows or columns without effecting the repairing bandwidth of $\bar{\MH}$. Hence,  WLOG, we can assume $\ell_1\le\ell_2$ and $\ell_3\le\ell_4$ for any $\bar{\MH}\in\CH\left\{4,2,(\ell_1,\ell_2,\ell_3,\ell_4)\right\}$. 
 By following the same trick in proving Lemma \ref{lm:optimalBH3}, we can have 
 \begin{IEEEeqnarray}{rCl}
    \sum_{r\in[4]}\ell_r|\CJ_r(\bar{\MH})|
    &=&\sum_{r\in[4]}\ell_r \left(\sum_{i\in[4]\setminus\{r\}}J^{(r)}_i(\bar{\MH})\right)\\
    &\le&\ell_1(\ell_2+1)+\ell_2(N-\ell_2)+\ell_3(\ell_4+1)+\ell_4(N-\ell_4),
 \end{IEEEeqnarray}
 as illustrated in Table \ref{tbl:JHtableforCH22}. 
 \begin{table}[ht]\centering
 \caption{ $J^{(r)}_i(\bar{\MH})$ for $\bar{\MH}\in\CH\left(4,2,(\ell_1,\ell_2,\ell_3,\ell_4)\right)$. }
 \label{tbl:JHtableforCH22}
  \begin{tabular}{|c|c|c|c|}\hline
   $\ell_1$&$\ell_2$&$\ell_3$&$\ell_4$\\\hline\hline
  \color{red}{$J^{(1)}_1(\bar{\MH})=0$} & $J^{(1)}_2(\bar{\MH})=\ell_2$ & $J^{(1)}_3(\bar{\MH})=1$ & $J^{(1)}_4(\bar{\MH})=0$\\\hline
   $J^{(2)}_1(\bar{\MH})=\ell_1$ &\color{red}{$J^{(2)}_2(\bar{\MH})=0$} & $J^{(2)}_3(\bar{\MH})=\ell_3$ & $J^{(2)}_4(\bar{\MH})=\ell_4$\\\hline
   $J^{(3)}_1(\bar{\MH})=0$ & $J^{(3)}_2(\bar{\MH})=1$ & \color{red}{$J^{(3)}_3(\bar{\MH})=0$} & $J^{(3)}_4(\bar{\MH})=\ell_4$\\\hline
   $J^{(4)}_1(\bar{\MH})=\ell_1$ & $J^{(4)}_2(\bar{\MH})=\ell_2$ & $J^{(4)}_3(\bar{\MH})=\ell_3$ & \color{red}{$J^{(4)}_4(\bar{\MH})=0$}\\\hline
  \end{tabular}
  \end{table}
  
  Therefore, we can rewrite \eqref{eq:BRLB} as
 \begin{multline}
  \bar{B}^*(R)\ge\min_{\vec{\ell}\in\{(\ell_1,\ell_2,\ell_3,\ell_4):\sum_{r\in[4]}\ell_r=N,\, \ell_1\le\ell_2,\,\ell_3\le\ell_4\}}
  2N(N-1)-\\\ell_1(\ell_2+1)-\ell_2(N-\ell_2)-\ell_3(\ell_4+1)-\ell_4(N-\ell_4)
  =\Delta_4,
 \end{multline}
 when $R_\textL\ge R-R_\textL\ge 2$. 
   
We  finally conclude that $\bar{B}^*(R)\ge\min\{\Delta_3,\Delta_4\}$ for all $R\in[N]$. 
\end{IEEEproof}

 \section{Numerical results}
\begin{figure}[ht]
  \centering
  \includegraphics[width=7in]{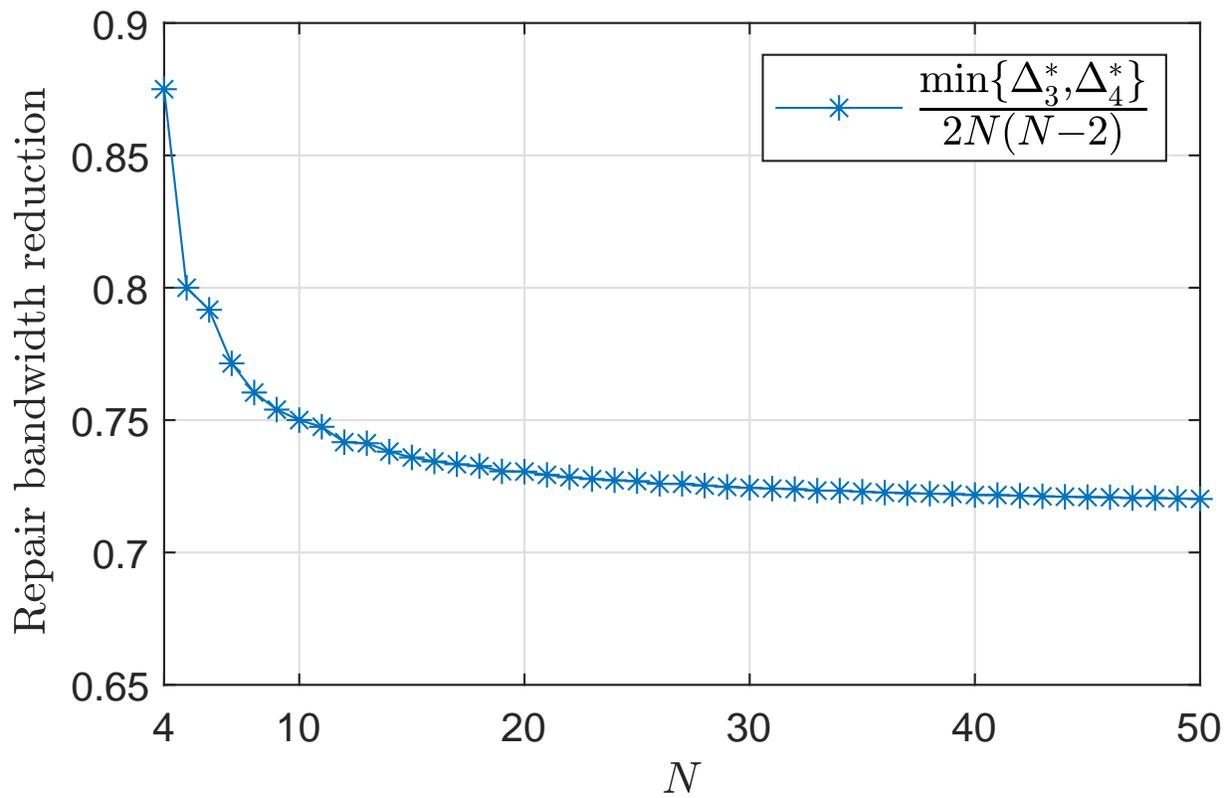}
  \caption{The lower bound $\min\{\Delta_3,\Delta_4\}$ from $N=4$ to $N=200$.}
  \label{fig:LB200}
 \end{figure}

 The derived lower bound, $\min\{\Delta_3,\Delta_4\}$ for $4\le N\le 200$, is plotted in Figure \ref{fig:LB200}. 
 The values at $N=26$ and $N=32$ are achieved by the codes provided by the department in Israel, 
 which implies that the derived bound is so tight that it can be achieved. 
 It should be noted that $\min\{\Delta_3,\Delta_4\}=\Delta_3$ for all $4\le N\le 200$, except when $N=4$ and $6$. 

\end{document}